\documentclass[english,aps,superscriptaddress,pre,showpacs,preprint,A4paper]{revtex4-1}
\usepackage[T1]{fontenc}
\usepackage[latin9]{inputenc}
\usepackage[active]{srcltx}
\usepackage{bm}
\usepackage{amsmath}
\usepackage{amssymb}
\usepackage{graphicx}
\usepackage{wasysym}

\makeatletter
\usepackage{subfigure}

\makeatother

\usepackage{babel}
\begin{document}

\title{Kinetic theory for a simple modeling of phase transition: Dynamics
out of local equilibrium}

\author{Shigeru Takata}

\affiliation{Department of Aeronautics and Astronautics, Graduate School of Engineering,
Kyoto University, Kyoto 615\textendash 8540, Japan}
\email{takata.shigeru.4a@kyoto-u.ac.jp}

\affiliation{Research Project of Fluid Science and Engineering, Advanced Engineering
Research Center, Kyoto University}

\author{Takuya Matsumoto}

\affiliation{Department of Aeronautics and Astronautics, Graduate School of Engineering,
Kyoto University, Kyoto 615\textendash 8540, Japan}

\author{Anna Hirahara}

\affiliation{Department of Aeronautics and Astronautics, Graduate School of Engineering,
Kyoto University, Kyoto 615\textendash 8540, Japan}

\author{Masanari Hattori}

\affiliation{Department of Aeronautics and Astronautics, Graduate School of Engineering,
Kyoto University, Kyoto 615\textendash 8540, Japan}

\affiliation{Research Project of Fluid Science and Engineering, Advanced Engineering
Research Center, Kyoto University}

\date{\today}
\begin{abstract}
This is a continuation of the previous work (Takata \& Noguchi, J.
Stat. Phys., 2018) that introduces the presumably simplest model of
kinetic theory for phase transition. Here, main concern is to clarify
the stability of uniform equilibrium states in the kinetic regime,
rather than that in the continuum limit. It is found by the linear
stability analysis that the linear neutral curve is invariant with
respect to the Knudsen number, though the transition process is dependent
on the Knudsen number. In addition, numerical computations of the (nonlinear) kinetic
model are performed to investigate the transition processes in detail.
Numerical results show that (unexpected) incomplete transitions
may happen as well as clear phase transitions.
\end{abstract}

\maketitle

\section{Introduction}

Recently, the first author has proposed a simple kinetic model for
the description of phase transition in Ref.~\cite{TN17}. In this
reference, the model is proposed as presumably the simplest kinetic
theory model that enables us to reproduce the phase transition phenomena.
A functional that decreases monotonically in time is also found for
this model. In the continuum limit~\cite{FN1} (or the local equilibrium), it
recovers the Cahn\textendash Hilliard type equation; based on this
limiting equation, the linear stability of uniform states is studied
to find a neutral curve and numerical computations of the Cahn\textendash Hilliard
type equation are carried out as well. The details of the phase transition
in the continuum limit have been clarified.

The model is, however, not limited to the dynamics in the continuum
limit in contrast to the lattice Boltzmann models (e.g. Refs.~\citep{SOOY96,GLS07}),
but rather aims at the dynamics in the kinetic regime (or out of local
equilibrium) as Refs.~\citep{G71,FGL05,KOW12,FB17}. In the present
paper, we take a further step to study the stability of uniform equilibrium
states in the kinetic regime. To be more specific, we are going to
study the stability problem by our kinetic theory model directly,
not through the Cahn\textendash Hilliard type equation, and try to
understand the influence of the Knudsen number. The paper is organized
as follows. After summarizing our model in Sec.~\ref{sec:Problem-and-its},
the dimensionless formulation is given in Sec.~\ref{sec:Dimensionless-notation}
for the clarity of similarity rule in the problem to be studied. Then,
a linear stability of uniform equilibrium states is investigated in
Sec.~\ref{sec:Linear-stability-of} in a way similar to Ref.~\cite{YP17},
but in more comprehensive way. The invariance of the neutral curve
with respect to the Knudsen number is shown as well. Results of numerical
computations, together with supplemental discussions, are presented
in Sec.~\ref{sec:Numerical-results-and}. The paper is concluded
in Sec.~\ref{sec:Conclusion}.

\section{Problem and its formulation\label{sec:Problem-and-its}}

Consider a system (or a fluid) composed of innumerable molecules in
a periodic spatial domain $D$:\begin{subequations}\label{model}
\begin{align}
\frac{\partial f}{\partial t}+ & \xi_{i}\frac{\partial f}{\partial X_{i}}+F_{i}\frac{\partial f}{\partial\xi_{i}}=C_{*}[f],\displaybreak[0]\label{eq:kinetic}\\
C_{*} & [f]=A(\rho)(\rho M_{*}-f),\ \rho=\int fd\bm{\xi},\displaybreak[0]\\
F_{i} & =-\frac{\partial\phi}{\partial X_{i}},\ \phi=\Phi_{S}+\Phi_{L},\displaybreak[0]\\
\Phi_{S} & =-RT_{*}\ln(1-b\rho)+\frac{b\rho RT_{*}}{1-b\rho}-2a\rho,\ (a,b>0),\displaybreak[0]\\
\Phi_{L} & =-\kappa\frac{\partial^{2}\rho}{\partial X_{i}^{2}},\ (\kappa>0),\displaybreak[0]\\
M_{*} & =\frac{1}{(2\pi RT_{*})^{3/2}}\exp(-\frac{\bm{\xi}^{2}}{2RT_{*}}),
\end{align}
\end{subequations}where $t$ is a time, $\bm{X}=(X_{1},X_{2},X_{3})$
a spatial position, $\bm{\xi}=(\xi_{1},\xi_{2},\xi_{3})$ a molecular
velocity, $f(t,\bm{X},\bm{\xi})$ a velocity distribution function
(VDF), $m\bm{F}=m(F_{1},F_{2},F_{3})$ a force acting on a single
molecule, with $m$ being its mass, and $\phi$ a corresponding potential.
The integration with respect to $\bm\xi$ (and its dimensionless counterpart $\bm\zeta$ that appears later)
is carried out over its entire space $\mathbb{R}^3$, unless otherwise stated.
$C_{*}[f]$ is the so-called collision term and plays a role of a
thermal bath, driving the system toward a thermal equilibrium at a
specified temperature $T_{*}$ which is fixed and given. Note that
the present collision term conserves neither momentum nor energy,
in contrast to the usual intermolecular collision term. $A$ is a
positive function of local density $\rho$ and $R=k_{B}/m$ with $k_{B}$
being the Boltzmann constant. $\kappa$, $a$, and $b$ are positive
constants, the latter two of which are the ones occurring in the van
der Waals equation of state {[}see (\ref{eq:VDW}) below{]}. Hence,
$\Phi_{S}$ is a potential purely related to the van der Waals equation
of state. In the meantime, $\Phi_{L}$ is a potential which comes
from a collection of nonlocal attractive interaction of molecules.
In the present paper, we exclusively consider the case where it can
be reduced to be local and of Laplacian form by the isotropic and
rapid decay assumptions in $\bm{X}$ (see Appendix~\ref{APP:0} and Ref.~\citep{TN17} for
details). We will investigate the time evolution of the system from
an initial distribution
\begin{equation}
f(0,\bm{X},\bm{\xi})=f_{\mathrm{in}}(\bm{X},\bm{\xi}),\label{eq:initial}
\end{equation}
a perturbed state from a uniform one $f=\rho_{0}M_{*}$ with $\rho_{0}$
being the initial average density of the fluid, aiming at studying
the stability of the uniform state. It is readily checked that the
mass is conserved in the present system. Accordingly, the average
density is constant in time.

The flow velocity $\bm{v}=(v_{1},v_{2},v_{3})$, stress tensor $p_{ij}$,
pressure $p$, and temperature $T$ are defined as\begin{subequations}\label{moments}
\begin{align}
\rho v_{i} & =\int\xi_{i}fd\bm{\xi},\ p_{ij}=\int c_{i}c_{j}fd\bm{\xi}+\int\rho\Phi_{S}^{'}(\rho)d\rho\delta_{ij},\displaybreak[0]\label{eq:moment1}\\
p & =\frac{1}{3}p_{ii},\ T=\frac{1}{3\rho R}\int\bm{c}^{2}fd\bm{\xi},\label{eq:moment2}
\end{align}
\end{subequations}where $c_{i}=\xi_{i}-v_{i}$ and $\prime$ denotes the derivative (with respect to $\rho$).
The above definition
of $p$ is consistent with the van der Waals equation of state
\begin{equation}
p=\frac{\rho RT}{1-b\rho}-\rho^{2}a,\label{eq:VDW}
\end{equation}
within the isothermal approximation $T\simeq T_{*}$. The non-isothermal
case is excluded in the present simple model, as discussed in detail
in Ref.~\cite{TN17}. For the present system, it is shown in Ref.~\cite{TN17}
that the following functional $\mathcal{M}$ monotonically decreases
in time:
\begin{align*}
\mathcal{M}(t)= & \int_{D}\{\int f\ln\frac{f}{\rho_{0}M_{*}}d\bm{\xi}+\int\frac{\Phi_{S}}{RT_{*}}d\rho+\frac{\rho\Phi_{L}}{2RT_{*}}\}d\bm{X}\displaybreak[0]\\
= & \int_{D}\{\int f\ln\frac{f}{\rho_{0}M_{*}}d\bm{\xi}+\int\frac{\Phi_{S}}{RT_{*}}d\rho\displaybreak[0]\\
 & \quad+\frac{\kappa}{2RT_{*}}(\frac{\partial\rho}{\partial X_{i}})^{2}\}d\bm{X}.
\end{align*}
Note that the second term of the above integrand is reduced
to
\[
\int\Phi_{S}d\rho=-\rho RT_{*}\ln(1-b\rho)-a\rho^{2},
\]
under the requirement that $\Phi_S$ vanishes in the low density limit $\rho\to0$.

To summarize, the above model describes a non-ideal gas that interacts
with a thermal bath through the simple collision term. Our concern is
the stability of uniform equilibrium states in such a closed non-isolated
system. Incidentally, we have recently become aware of Ref.~\citep{BB16}
which studies the stability of uniform equilibrium states by the kinetic
theory.  However, what is considered there is an ideal gas under a collective
attractive interaction in an isolated system, and accordingly the
studied phenomenon is qualitatively different. Indeed, the details
such as the linear stability analysis and its results, the derived
H theorem, etc. are all different from ours.

\section{Dimensionless notation\label{sec:Dimensionless-notation}}

For the later convenience and for the clarity of similarity law for
the present system, we introduce the following dimensionless notation:\begin{subequations}\label{factor}
\begin{align}
 & t=\frac{L}{(2RT_{*})^{1/2}}\tilde{t},\ X_{i}=Lx_{i},\ \xi_{i}=(2RT_{*})^{1/2}\zeta_{i},\displaybreak[0]\label{eq:dless1}\\
 & \rho=\rho_{0}\tilde{\rho,}\ f=\frac{\rho_{0}}{(2RT_{*})^{3/2}}\tilde{f},\ A(\rho)=A(\rho_{0})\tilde{A}(\tilde{\rho}),\displaybreak[0]\\
 & \phi=2RT_{*}\tilde{\phi},\ \Phi_{S}=2RT_{*}\tilde{\Phi}_{S},\ \Phi_{L}=2RT_{*}\tilde{\Phi}_{L},\displaybreak[0]\\
 & \kappa=\frac{2RT_{*}L^{2}}{\rho_{0}}\tilde{\kappa},\ a=\frac{\tilde{a}RT_{*}}{\rho_{0}},\ b=\frac{\tilde{b}}{\rho_{0}},
\end{align}
\end{subequations}where $\bm{x}=(x_{1},x_{2},x_{3})$, $\bm{\zeta}=(\zeta_{1},\zeta_{2},\zeta_{3})$,
$\zeta=|\bm{\zeta}|$, and $L$ is the characteristic length of the
system, typically its period. Then the original equation and the initial
condition are recast as\begin{subequations}\label{modeldless}
\begin{align}
\frac{\partial\tilde{f}}{\partial\tilde{t}} & +\zeta_{i}\frac{\partial\tilde{f}}{\partial x_{i}}-\frac{\partial\tilde{\phi}}{\partial x_{i}}\frac{\partial\tilde{f}}{\partial\zeta_{i}}=\frac{2}{\sqrt{\pi}}\frac{\tilde{A}(\tilde{\rho})}{\mathrm{Kn}}(\tilde{\rho}E-\tilde{f}),\label{eq:dimensionlessKM}\\
 & \tilde{f}(0,\bm{x},\bm{\zeta})=\tilde{f}_{\mathrm{in}}(\bm{x},\bm{\zeta}),
\end{align}
\end{subequations}where\begin{subequations}
\begin{align}
 & \tilde{\rho}=\int\tilde{f}d\bm{\zeta},\ E=\pi^{-3/2}\exp(-\zeta^{2}),\ \tilde{\phi}=\tilde{\Phi}_{S}+\tilde{\Phi}_{L},\displaybreak[0]\\
 & \tilde{\Phi}_{S}(\tilde{\rho})=-\frac{1}{2}\ln(1-\tilde{b}\tilde{\rho})-\tilde{a}\tilde{\rho}+\frac{1}{2}\frac{\tilde{b}\tilde{\rho}}{1-\tilde{b}\tilde{\rho}},\ (\tilde{a},\tilde{b}>0),\label{eq:defPhi}\\
 & \tilde{\Phi}_{L}=-\tilde{\kappa}\frac{\partial^{2}\tilde{\rho}}{\partial x_{i}^{2}},\ (\tilde{\kappa}>0),\qquad\tilde{A}(\tilde{\rho})>0,
\end{align}
and $\mathrm{Kn}$ is the Knudsen number defined as
\begin{equation}
\mathrm{Kn}=\frac{(8RT_{*}/\pi)^{1/2}}{A(\rho_{0})L}.
\end{equation}
\end{subequations}In the above, the Strouhal number is set to be
unity by the specific choice of reference time $L/(2RT_{*})^{1/2}$
in (\ref{eq:dless1}).

The moments of $f$, i.e., $v_{i}=(2RT_{*})^{1/2}\tilde{v}_{i}$,
$p_{ij}=\rho_{0}RT_{*}\tilde{p}_{ij}$, $p=\rho_{0}RT_{*}\tilde{p}$,
and $T=T_{*}\tilde{T}$ are recast as moments of $\tilde{f}$:\begin{subequations}\label{macrodless}
\begin{align}
\tilde{\rho}\tilde{v}_{i} & =\int\zeta_{i}\tilde{f}d\bm{\zeta},\ \tilde{p}_{ij}=2\!\int\tilde{c}_{i}\tilde{c}_{j}\tilde{f}d\bm{\zeta}+2\!\int\tilde{\rho}\tilde{\Phi}_{S}^{\prime}d\tilde{\rho}\,\delta_{ij},\displaybreak[0]\\
\tilde{p} & =\frac{1}{3}\tilde{p}_{ii},\ \tilde{T}=\frac{2}{3\tilde{\rho}}\int\tilde{\bm{c}}^{2}\tilde{f}d\bm{\zeta},
\end{align}
\end{subequations}where $\tilde{\bm{c}}=\bm{\zeta}-\tilde{\bm{v}}$.
Furthermore, the equation of state (\ref{eq:VDW}) and the monotonically
decreasing functional $\mathcal{M}=\rho_{0}L^{3}\tilde{\mathcal{M}}$
are rewritten as
\begin{align}
\tilde{p} & =\frac{\tilde{\rho}\tilde{T}}{1-\tilde{b}\tilde{\rho}}-\tilde{a}\tilde{\rho}^{2},\displaybreak[0]\label{eq:vdw_dless}\\
\tilde{\mathcal{M}}(\tilde{t}) & =\int_{\tilde{D}}\{\int\tilde{f}\ln\frac{\tilde{f}}{E}d\bm{\zeta}+2\int\tilde{\Phi}_{S}d\tilde{\rho}+\tilde{\rho}\tilde{\Phi}_{L}\}d\bm{x}\nonumber \\
 & =\int_{\tilde{D}}\{\int\tilde{f}\ln\frac{\tilde{f}}{E}d\bm{\zeta}+2\int\tilde{\Phi}_{S}d\tilde{\rho}-\tilde{\kappa}\tilde{\rho}\frac{\partial^{2}\tilde{\rho}}{\partial x_{i}^{2}}\}d\bm{x}\nonumber \\
 & =\int_{\tilde{D}}\{\int\tilde{f}\ln\frac{\tilde{f}}{E}d\bm{\zeta}+2\int\tilde{\Phi}_{S}d\tilde{\rho}+\tilde{\kappa}\Big(\frac{\partial\tilde{\rho}}{\partial x_{i}}\Big)^{2}\}d\bm{x},\label{eq:Mdless}
\end{align}
where $\tilde{D}$ is the counterpart of $D$ and
\[
2\int\tilde{\Phi}_{S}d\tilde{\rho}=-\tilde{a}\tilde{\rho}^{2}-\tilde{\rho}\ln(1-\tilde{b}\tilde{\rho}).
\]
Note that the dimensionless mass in the domain $\tilde{D}$ is invariant
in time and the average dimensionless density is unity, i.e.,
\[
\frac{1}{V_{\tilde{D}}}\int_{\tilde{D}}\tilde{\rho}\,d\bm{x}=1,
\]
where $V_{\tilde{D}}$ is the volume of $\tilde{D}$.

\section{Linear stability of uniform equilibrium states\label{sec:Linear-stability-of}}

In the present section, we will study the linear stability of the
uniform equilibrium state $\tilde{f}=E$. To this end, we first substitute
$\tilde{f}=E+g$ into (\ref{eq:dimensionlessKM}) and then retain
only the linear terms in $g$, assuming $|g|\ll E$, to have
\begin{equation}
\alpha\frac{\partial g}{\partial\tilde{t}}+\alpha\zeta_{i}\frac{\partial g}{\partial x_{i}}-2\alpha(\tilde{\kappa}\frac{\partial^{2}}{\partial x_{j}^{2}}-\tilde{\Phi}_{S}^{\prime}(1))\zeta_{i}\frac{\partial\tilde{\rho}_{g}}{\partial x_{i}}E=\tilde{\rho}_{g}E-g,\label{eq:gsys}
\end{equation}
where $\prime$ denotes the derivative (with respect to $\tilde{\rho}$)
and
\[
\alpha=\frac{\sqrt{\pi}}{2}\mathrm{Kn}(>0),\quad\tilde{\rho}_{g}=\int g\,d\bm{\zeta}.
\]
Now we are going to study whether the perturbation of the form $g=\exp(\sigma\tilde{t}+i\bm{\lambda}\cdot\bm{x})h(\bm{\zeta})$,
where $\bm{\lambda}=(\lambda_{1},\lambda_{2},\lambda_{3})$ with $\lambda_{i}$
being a positive number and $\sigma\in\mathbb{C}$, grows (i.e., $\Re\sigma>0$)
or decays (i.e., $\Re\sigma<0$) in time, thereby finding the stability
condition of the linearized system (\ref{eq:gsys}). Substituting
it into (\ref{eq:gsys}) eventually yields the following identity
for $h$:\citep{YP17}
\begin{align}
h= & \frac{1}{\alpha\lambda}\Big(\frac{S-P(\bm{\zeta}\cdot\bm{e})\alpha\lambda[Q+(\bm{\zeta}\cdot\bm{e})]}{S^{2}+[Q+(\bm{\zeta}\cdot\bm{e})]^{2}}\nonumber \\
 & \quad-\frac{i\{P(\bm{\zeta}\cdot\bm{e})S\alpha\lambda+[Q+(\bm{\zeta}\cdot\bm{e})]\}}{S^{2}+[Q+(\bm{\zeta}\cdot\bm{e})]^{2}}\Big)\tilde{\rho}_{h}E,\label{eq:h}
\end{align}
where $\tilde{\rho}_{h}=\int hd\bm{\zeta}$, $\sigma=\sigma_{1}+i\sigma_{2}$
($\sigma_{1},\sigma_{2}\in\mathbb{R}$), $\bm{e}=\bm{\lambda}/\lambda$,
$\lambda=|\bm{\lambda}|$, and
\begin{equation}
S=\frac{1+\alpha\sigma_{1}}{\alpha\lambda},\ P=2[\tilde{\kappa}\lambda^{2}+\tilde{\Phi}_{S}^{\prime}(1)],\ Q=\frac{\sigma_{2}}{\lambda}.\label{eq:defSPQ}
\end{equation}
Due to the consistency, integrating both sides of (\ref{eq:h}) in
$\bm{\zeta}$ leads to the following set of identities:
\begin{align}
 & 2\pi\int_{0}^{\infty}\int_{0}^{\pi}\frac{1}{\alpha\lambda}\frac{S-(\zeta\cos\theta)P\bar{Q}\alpha\lambda}{S^{2}+\bar{Q}^{2}}E\,\zeta^{2}\sin\theta d\theta d\zeta=1,\displaybreak[0]\label{eq:cond1-1}\\
 & \int_{0}^{\infty}\int_{0}^{\pi}\frac{PS\alpha\lambda\zeta\cos\theta+\bar{Q}}{S^{2}+\bar{Q}^{2}}E\zeta^{2}\sin\theta d\theta d\zeta=0,\label{eq:cond2-1}
\end{align}
where $\bar{Q}\equiv Q+\zeta\cos\theta$ and $\theta$ is the angle
between $\bm{\zeta}$ and $\bm{e}$, i.e., $\bm{\zeta}\cdot\bm{e}=\zeta\cos\theta$.
After some manipulations, (\ref{eq:cond1-1}) is transformed into
\begin{equation}
(1+PS\alpha\lambda)SI+P\alpha\lambda Q^{2}(I-4J)=\frac{\alpha\lambda}{\pi}(1+P),\label{eq:cond1red}
\end{equation}
while (\ref{eq:cond2-1}) is transformed into
\begin{equation}
Q\{(1+PS\alpha\lambda)(I-4J)-PS\alpha\lambda I\}=0,\label{eq:cond2red}
\end{equation}
where both $I$ and $J$ are positive functions of $Q$ and $S$ defined
by
\begin{align*}
I & (Q,S)=\int_{0}^{\infty}(\frac{1}{S^{2}+(Q+\zeta)^{2}}+\frac{1}{S^{2}+(Q-\zeta)^{2}})Ed\zeta,\displaybreak[0]\\
J & (Q,S)=\int_{0}^{\infty}\frac{\zeta^{2}E}{\{S^{2}+(Q+\zeta)^{2}\}\{S^{2}+(Q-\zeta)^{2}\}}d\zeta.
\end{align*}
Note that both $I$ and $J$ are even in $Q$.

\paragraph*{Case 1}

When $Q=0$, (\ref{eq:cond2red}) is automatically satisfied. Then,
(\ref{eq:cond1red}) with $Q=0$ takes the form
\[
2S(1+PS\alpha\lambda)\int_{0}^{\infty}\frac{E}{S^{2}+\zeta^{2}}d\zeta=\frac{\alpha\lambda}{\pi}(1+P),
\]
which is solved for $P$ and is reduced to
\begin{align*}
 & P=\frac{\alpha\lambda-\sqrt{\pi}F(S)}{\sqrt{\pi}\alpha\lambda SF(S)-\alpha\lambda},\\
 & F(S)\equiv2\sqrt{\pi}S\int_{0}^{\infty}\frac{E}{S^{2}+\zeta^{2}}d\zeta=\exp(S^{2})\{1-\mathrm{erf}(S)\},
\end{align*}
where $\mathrm{erf}(x)=(2/\sqrt{\pi})\int_{0}^{x}\exp(-s^{2})ds$.
Using the definitions of $P$ and $\tilde{\Phi}_{S}$ {[}see (\ref{eq:defSPQ})
and (\ref{eq:defPhi}){]}, we obtain \begin{subequations}\label{eq:cond_a}
\begin{align}
\tilde{a}= & \tilde{\kappa}\lambda^{2}+\frac{1}{2}\frac{\tilde{b}(2-\tilde{b})}{(1-\tilde{b})^{2}}+\frac{1}{2}X,\displaybreak[0]\label{eq:neutral}\\
 & X(S,\alpha\lambda)=\frac{\alpha\lambda-\sqrt{\pi}F(S)}{\alpha\lambda\{1-\sqrt{\pi}SF(S)\}}.\label{eq:X}
\end{align}
\end{subequations}As seen from its definition, $F(S)$ is monotonically
decreasing in $S$ from unity to zero and $x>\sqrt{\pi}F(1/x)$ for
$x>0$, while $SF(S)$ is monotonically increasing from zero to $1/\sqrt{\pi}$
in the range $0\le S<\infty$ \citep{Note1}. Hence, for $S\alpha\lambda>1$,
which is equivalent to $\sigma_{1}>0$, $X$ is monotonically increasing
in $S$ and approaches its infimum when $S\to1/(\alpha\lambda)$:
\[
X(S,\alpha\lambda)>X(\frac{1}{\alpha\lambda},\alpha\lambda)=\frac{\alpha\lambda-\sqrt{\pi}F(1/(\alpha\lambda))}{\alpha\lambda-\sqrt{\pi}F(1/(\alpha\lambda))}=1.
\]
Therefore, when $Q=0$, the uniform equilibrium state is unstable
if
\[
\tilde{a}>\frac{1}{2}\frac{\tilde{b}(2-\tilde{b})}{(1-\tilde{b})^{2}}+\frac{1}{2}=\frac{1}{2(1-\tilde{b})^{2}}.
\]

\paragraph*{Case 2}

When $Q\ne0$, the following two conditions must be satisfied simultaneously:\begin{subequations}
\begin{align}
 & (1+PS\alpha\lambda)SI+P\alpha\lambda Q^{2}(I-4J)=\frac{\alpha\lambda}{\pi}(1+P),\displaybreak[0]\\
 & (1+PS\alpha\lambda)(I-4J)-PS\alpha\lambda I=0.
\end{align}
\end{subequations}
In this case, using some properties of the Voigt functions \cite{NIST},
we can show that there is no mode that grows exponentially in time,
so that uniform equilibrium states are stable for $Q\ne0$.
See Appendix~\ref{APP:A} for the details of analysis.
\begin{figure}
\centering
\subfigure[]{\includegraphics[bb=0bp 0bp 448bp 312bp,clip,scale=0.4]{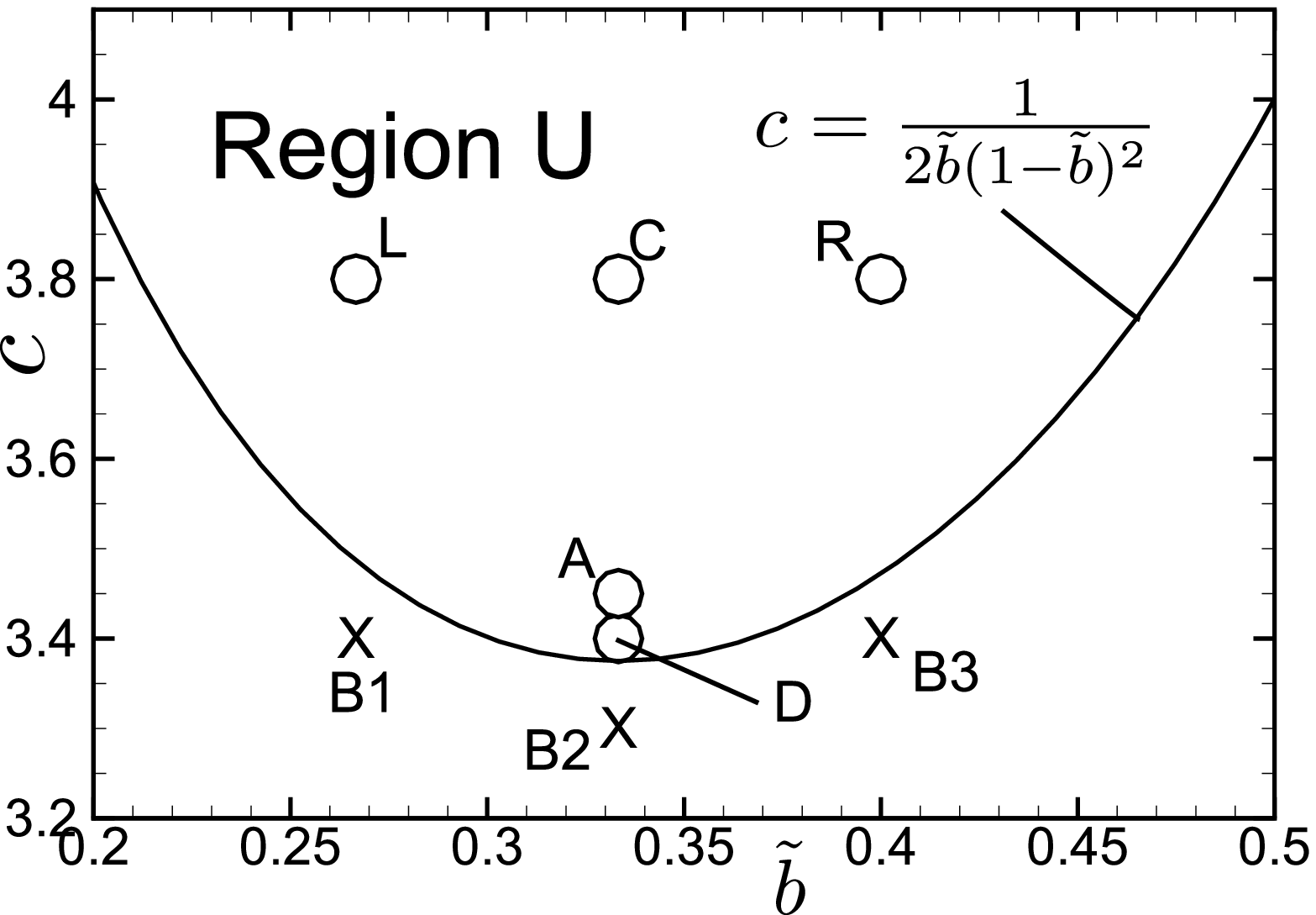}}\qquad
\subfigure[]{\includegraphics[bb=0bp 0bp 444bp 310bp,clip,scale=0.4]{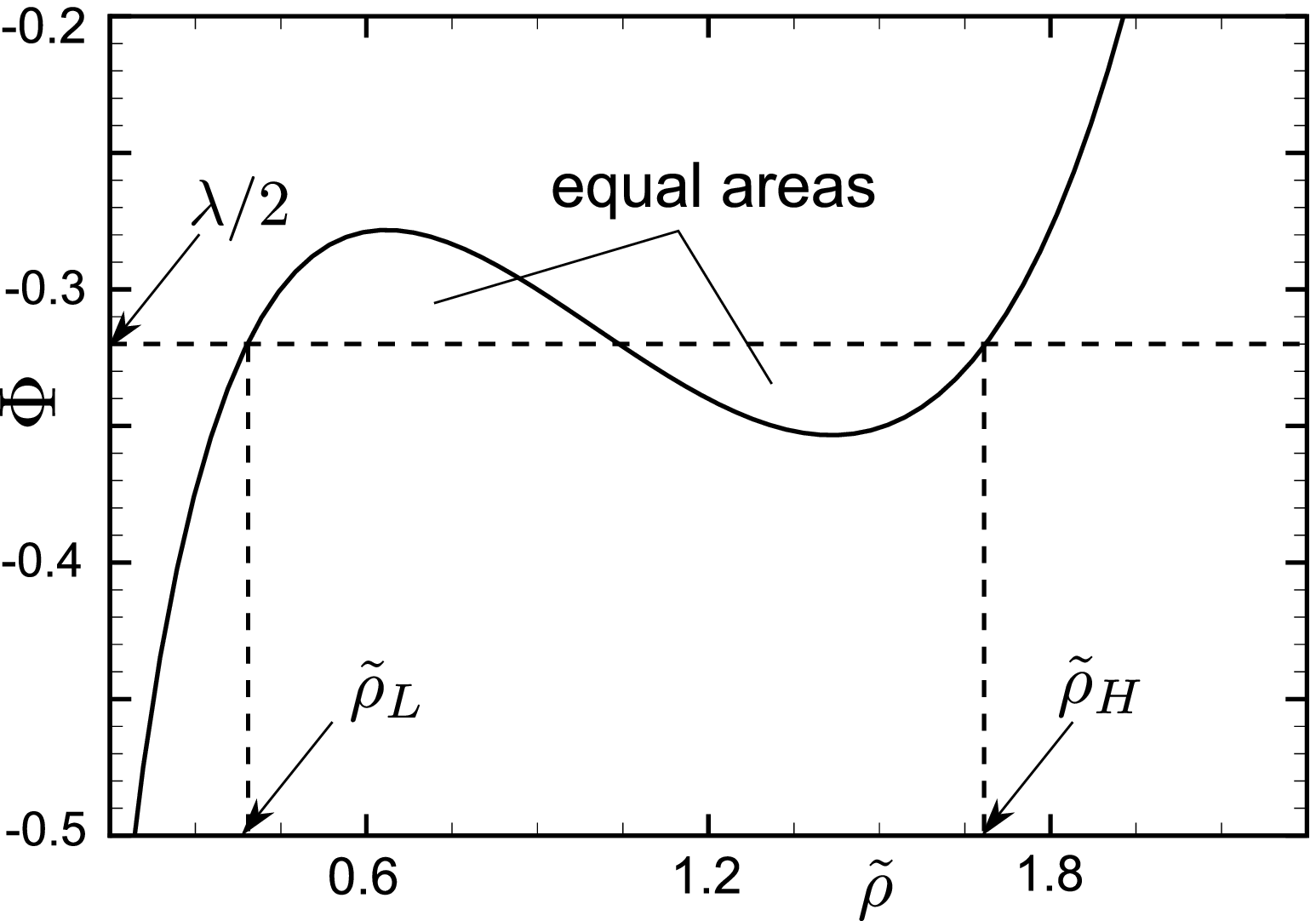}}
\caption{Diagrams of phase transition. (a) Neutral curve of the linear stability.
Markers with L, C, R, D, and A in the region U indicate the cases
$(c,\tilde{b})=(3.8,4/15)$, (3.8,$1/3)$, $(3.8,2/5)$, $(3.4,1/3)$,
and $(3.45,1/3)$, respectively. In the region U above the
neutral curve $c=1/\{2\tilde{b}(1-\tilde{b})^{2}\}$, uniform equilibrium
states are linearly unstable. Markers with B1, B2, and B3 indicate
the case $(c,\tilde{b})=(3.4,4/15)$, $(3.3,1/3)$, and $(3.4,2/5)$,
respectively. $\Circle$ ($\times$) indicates that phase transition
is observed (not observed) in numerical computations. Note that $\tilde{b}$
is a measure of the volume fraction of molecules to the total volume.
(b) Equiarea rule for determining two densities in stationary states
after phase transition. The curve is the one for case C in (a), i.e.,
$(c,\tilde{b})=(3.8,1/3)$.\label{fig:diagram}}
\end{figure}

In the long run, we conclude that there is a mode that grows exponentially
in time if the following condition is satisfied {[}see Fig.~\ref{fig:diagram}(a){]}:
\begin{equation}
c\equiv\frac{\tilde{a}}{\tilde{b}}>\frac{1}{2\tilde{b}(1-\tilde{b})^{2}}.\label{eq:instability condition}
\end{equation}
Note that this coincides with that obtained from the linear stability
analysis based on the Cahn\textendash Hilliard type equation (see
Ref.~\cite{TN17}; $\chi_{\mathrm{av}}$ in this reference is identical
to the present $\tilde{b}$). The linear instability condition is,
thus, invariant with respect to the Knudsen number, including the
continuum limit.

In closing the present section, we make a remark
that the substitution of the specific form of $g$ is a standard way in the linear stability analysis \cite{C81}
and is identical to taking the Laplace and Fourier transforms, in time and space respectively, of equation~\eqref{eq:gsys}.
Since the system is linear, the principle of superposition applies;
the obtained results are general accordingly.
One missing point that should have been taken into account is the period of spatial domain.
It does not admit the growing modes with a period longer than itself,
which possibly modifies the linear neutral curve. \cite{FNN}
We shall come back to this point later together with numerical evidences and
supplemental discussions at the end of Sec.~\ref{sec:Numerical-results-and}.

\section{Numerical results and discussions\label{sec:Numerical-results-and}}

\begin{figure}
\centering
\subfigure[]{\includegraphics[clip,scale=0.4]{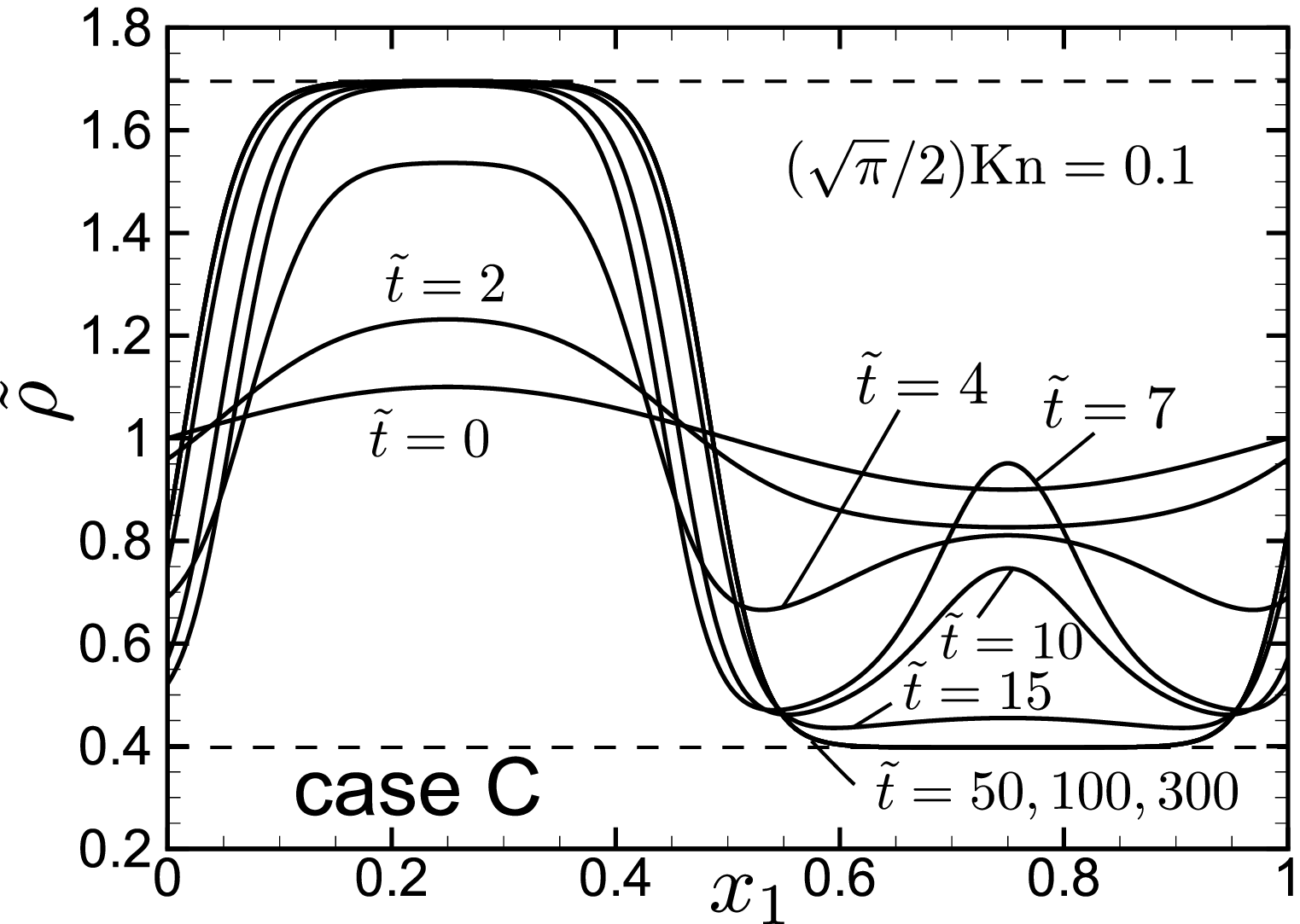}}\qquad
\subfigure[]{\includegraphics[clip,scale=0.4]{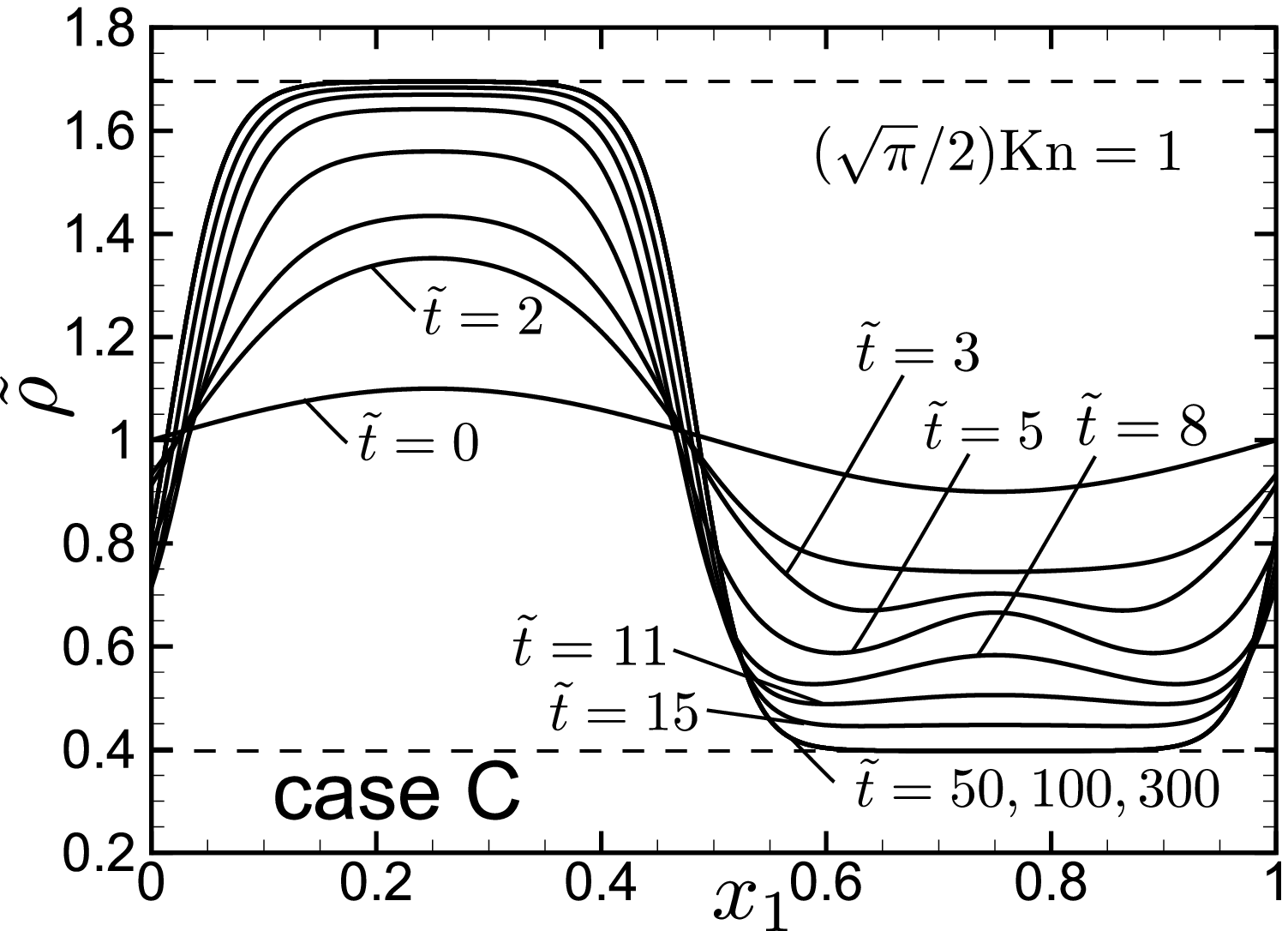}}\\
\subfigure[]{\includegraphics[clip,scale=0.4]{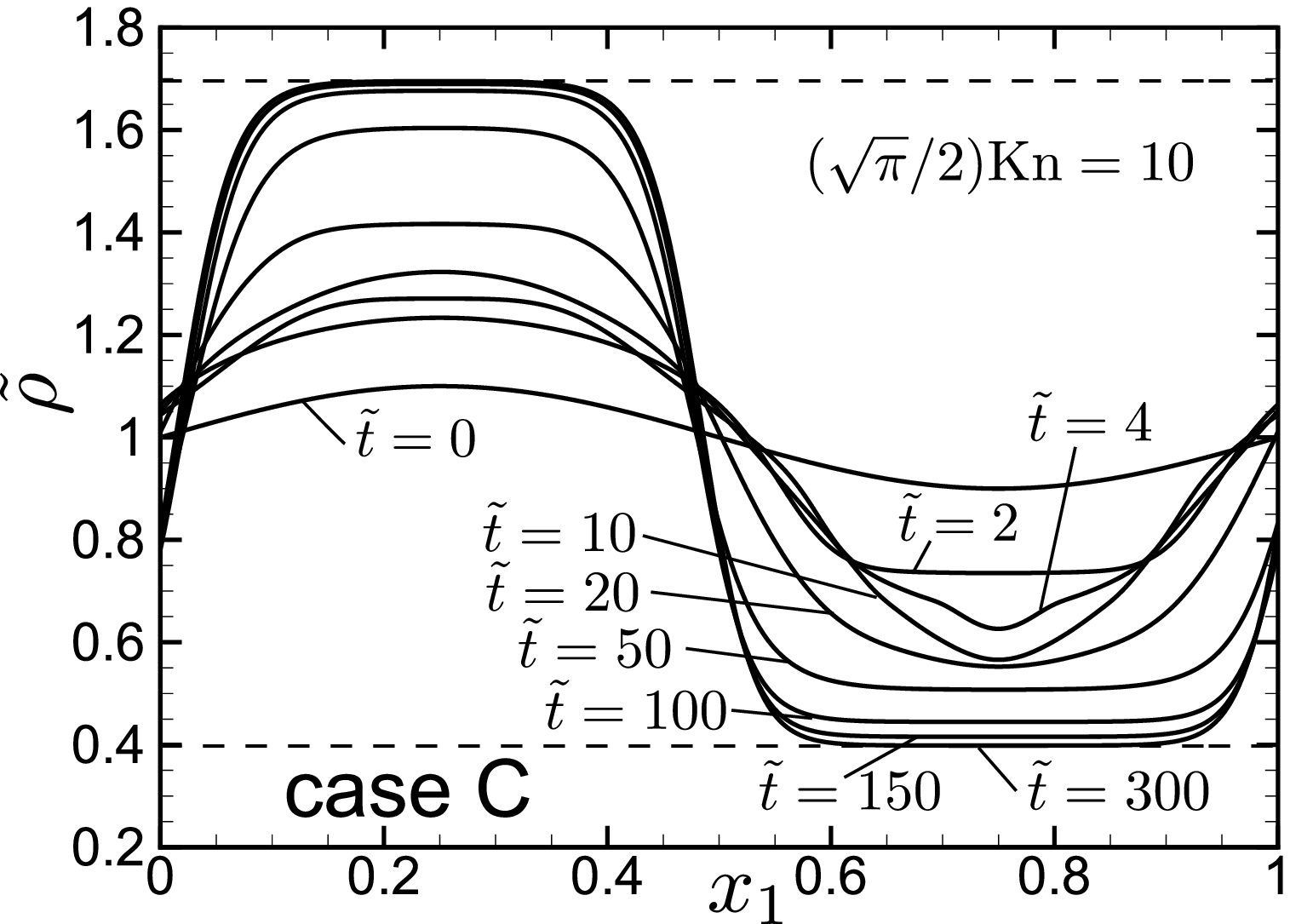}}\qquad
\subfigure[]{\includegraphics[bb=0bp 0bp 437bp 311bp,clip,scale=0.4]{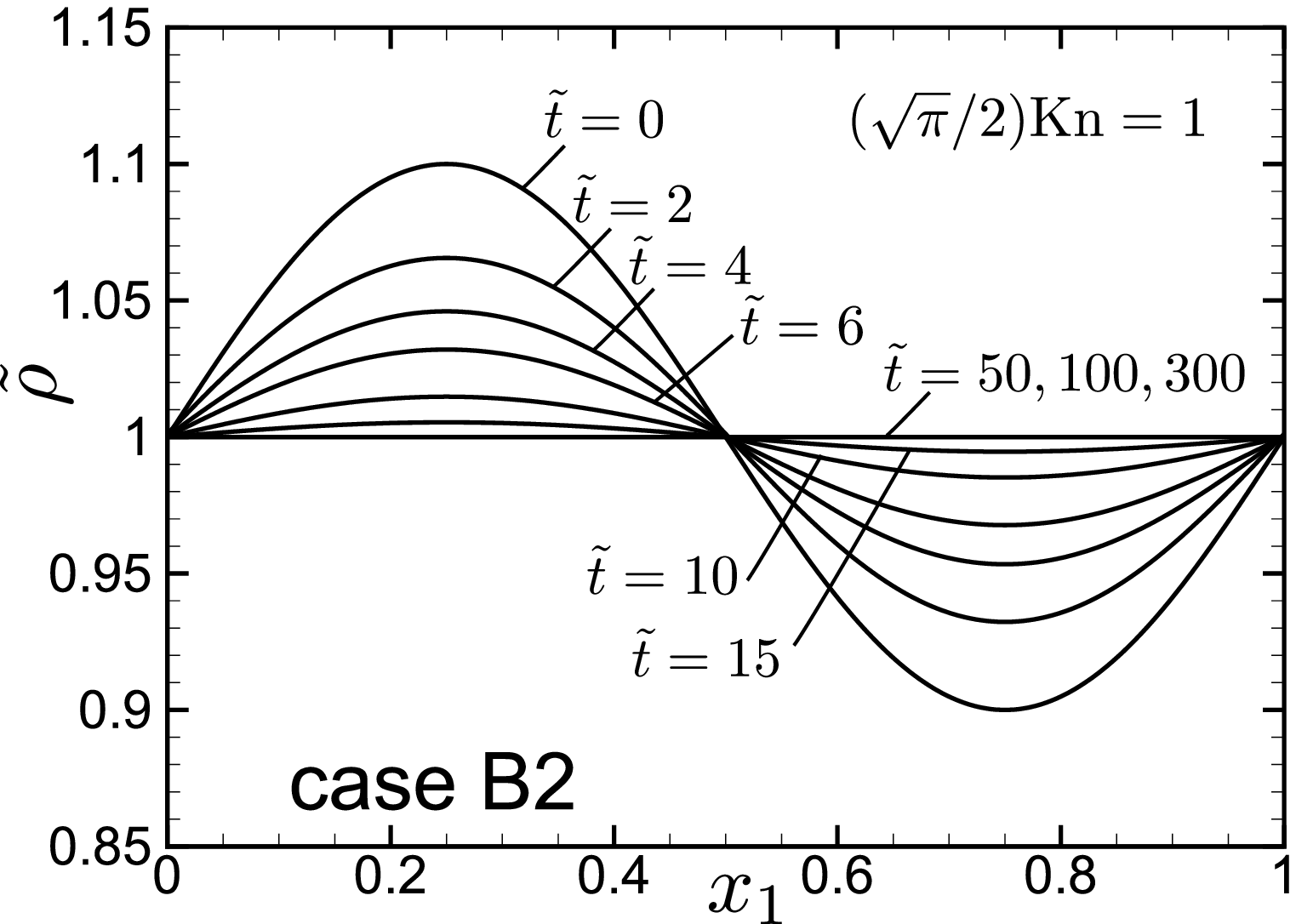}}

\caption{Time evolution of the density profile for cases C and B2 with $\tilde{\kappa}/\tilde{b}=5\times10^{-4}$.
(a) $(\sqrt{\pi}/2)\mathrm{Kn}=0.1$, (b) $(\sqrt{\pi}/2)\mathrm{Kn}=1$, and
(c) $(\sqrt{\pi}/2)\mathrm{Kn}=10$ for case C, while (d) $(\sqrt{\pi}/2)\mathrm{Kn}=1$
for case B2. Top and bottom dashed lines in panels (a)--(c) indicate $\tilde{\rho}_{H}$
and $\tilde{\rho}_{L}$ predicted by the equiarea rule in Fig.~\ref{fig:diagram}(b).
\label{fig:Time-evolution-of}}
\end{figure}
In the present section, we present some results of numerical computations
of the dimensionless system (\ref{eq:dimensionlessKM}) for spatially
one-dimensional case in the domain $\tilde{D}=\{x_{1}|0\le x_{1}<1\}$
with the initial condition
\[
\tilde{f}_{\mathrm{in}}(x_{1},\bm{\zeta})=\{1+\epsilon\sin(2\pi x_{1})\}E,
\]
where $\epsilon=0.1$. Because the form of $\tilde{f}=\psi(\tilde{t},x_{1},\zeta_{1})\pi^{-1}\exp(-\zeta_{2}^{2}-\zeta_{3}^{2})$
is compatible to the above system and $\tilde{\rho}$ can be computed
from $\psi$, the problem can be reduced to that of $\psi$, meaning
a great reduction of computational cost. Once $\psi$ is known, the
functional $\tilde{\mathcal{M}}$ can be recovered as well, because
$\int\tilde{f}\ln(\tilde{f}/E)d\bm{\zeta}=\int\psi\ln(\psi/E_{1})d\zeta_{1}$
with $E_{1}=\pi^{-1/2}\exp(-\zeta_{1}^{2})$. In the actual computation,
we adopt a semi-Lagrangian method (see, e.g., Refs.~\cite{QC10,EO14,GRS14})
based on the Strang's splitting \citep{Note2} with uniform grids
both in $x_{1}$ and $\zeta_{1}$, where the infinite domain of $\zeta_{1}$
is truncated into $|\zeta_{1}|\le6$. In each transport processes,
the 2nd\textendash 3rd WENO interpolation has been used. The second-order
central finite-difference is repeatedly applied to approximate the
third-order derivative of $\tilde{\rho}$ occurring in the gradient
of $\tilde{\Phi}_{L}$. We have also developed a finite-difference
scheme and a Strang's scheme with third-order polynomial spline interpolation
in place of the WENO interpolation. Different methods gave consistent
results as increasing grid points. We omit further details on the
numerical method itself and proceed to the presentation of results.
\begin{figure}
\centering\subfigure[]{\includegraphics[clip,scale=0.4]{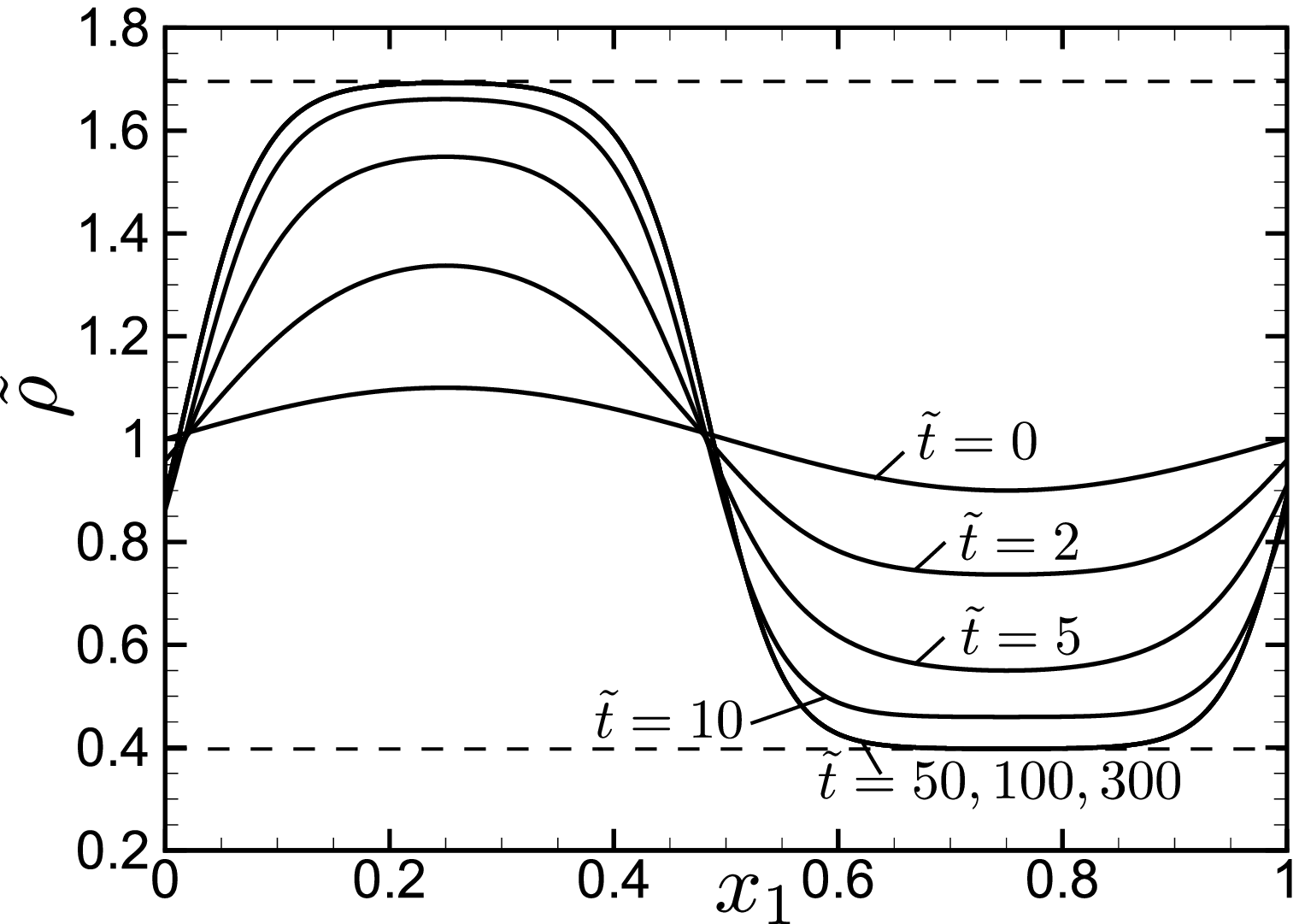}}\qquad
\subfigure[]{\includegraphics[clip,scale=0.4]{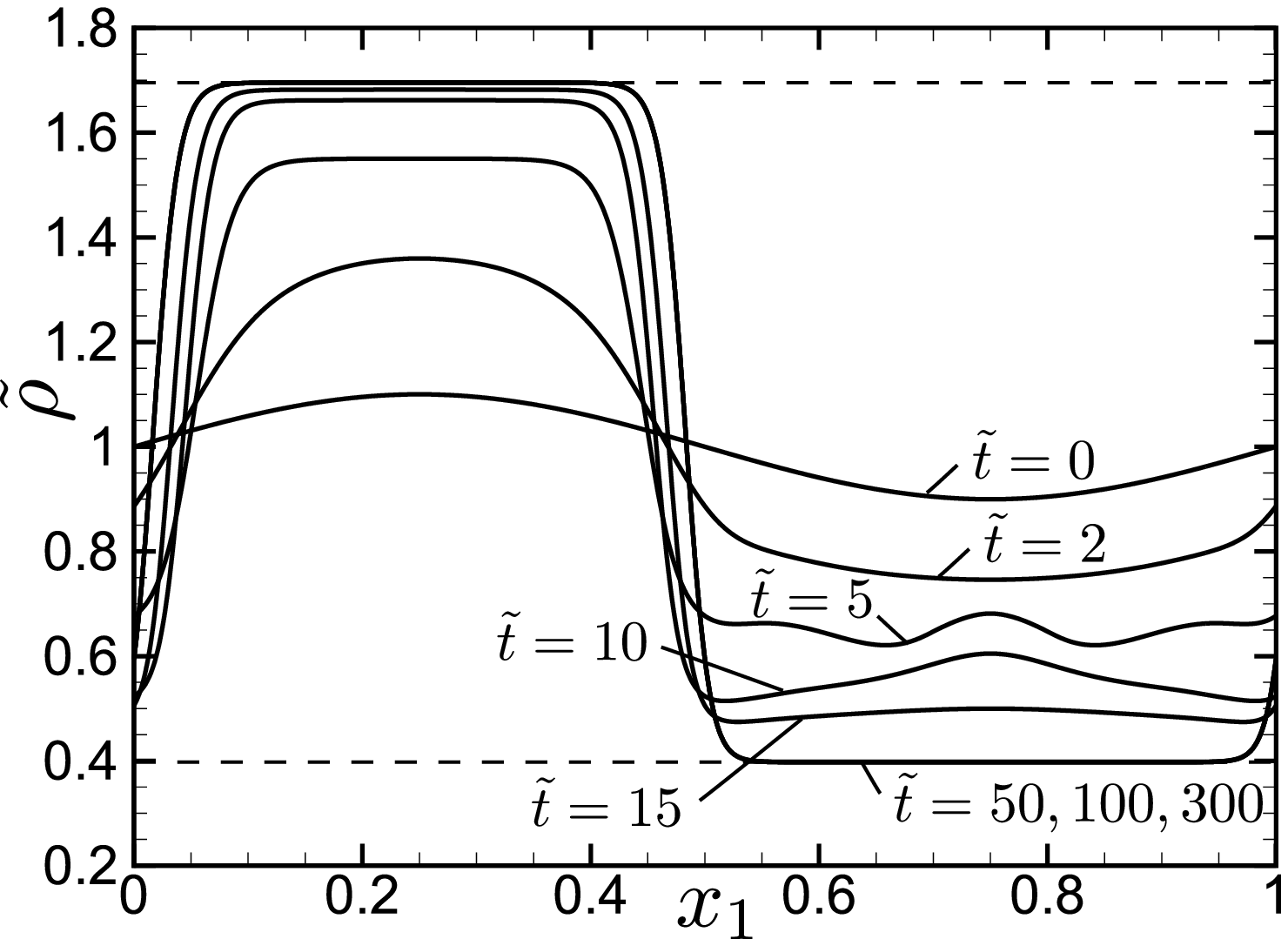}}

\caption{Time evolution of the density profile for case C
for $(\sqrt{\pi}/2)\mathrm{Kn}=1$. (a) $\tilde{\kappa}/\tilde{b}=1\times10^{-3}$
and (b) $\tilde{\kappa}/\tilde{b}=1\times10^{-4}$. See also Fig.~\ref{fig:Time-evolution-of}(b)
for comparison. Top and bottom dashed lines in each panel indicate
$\tilde{\rho}_{H}$ and $\tilde{\rho}_{L}$ predicted by the equiarea
rule in Fig.~\ref{fig:diagram}(b). \label{fig:Time-evolution-of-1}}
\end{figure}
\begin{figure}
\centering\subfigure[]{\includegraphics[clip,scale=0.4]{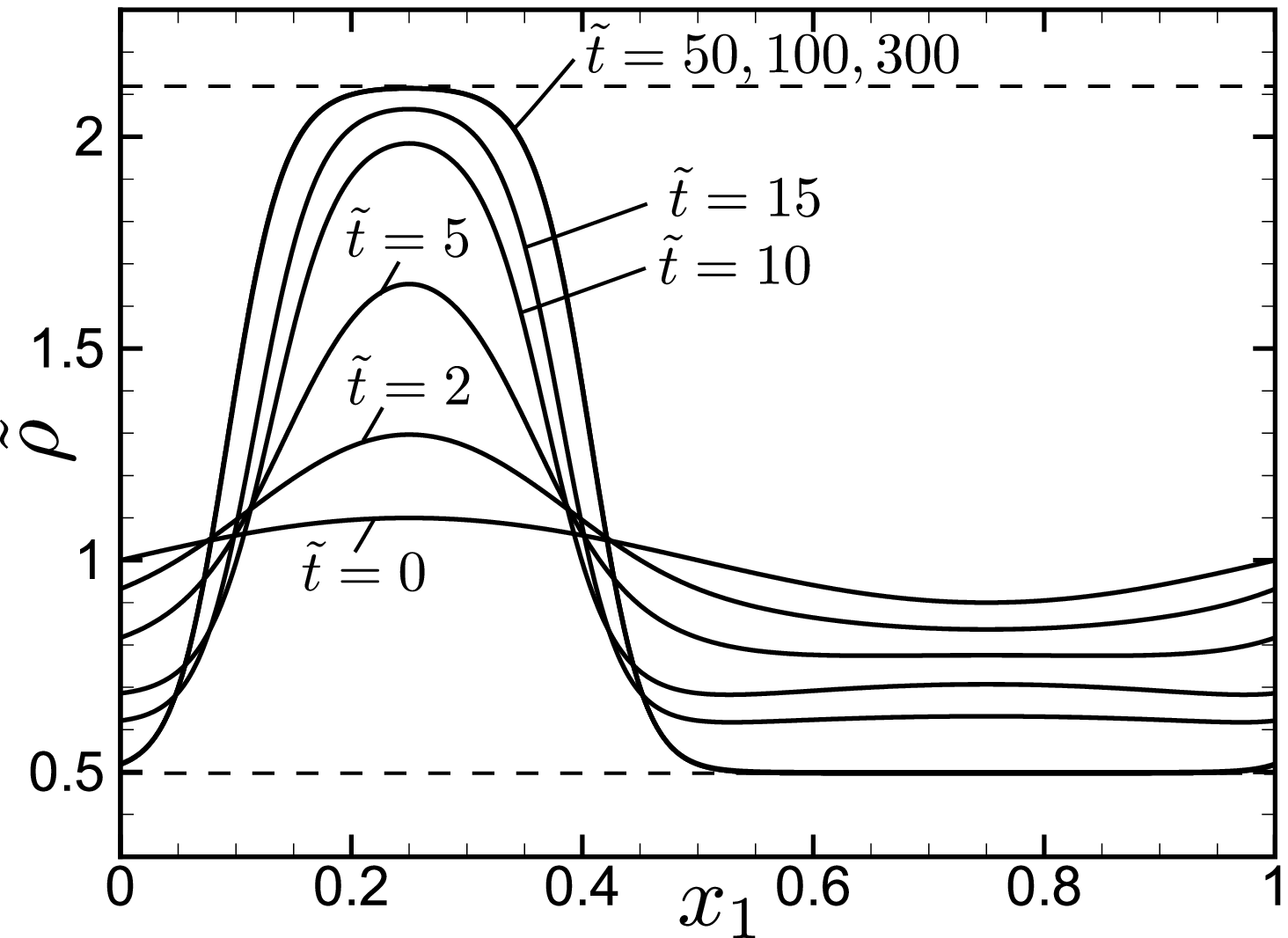}}\qquad\subfigure[]{\includegraphics[clip,scale=0.4]{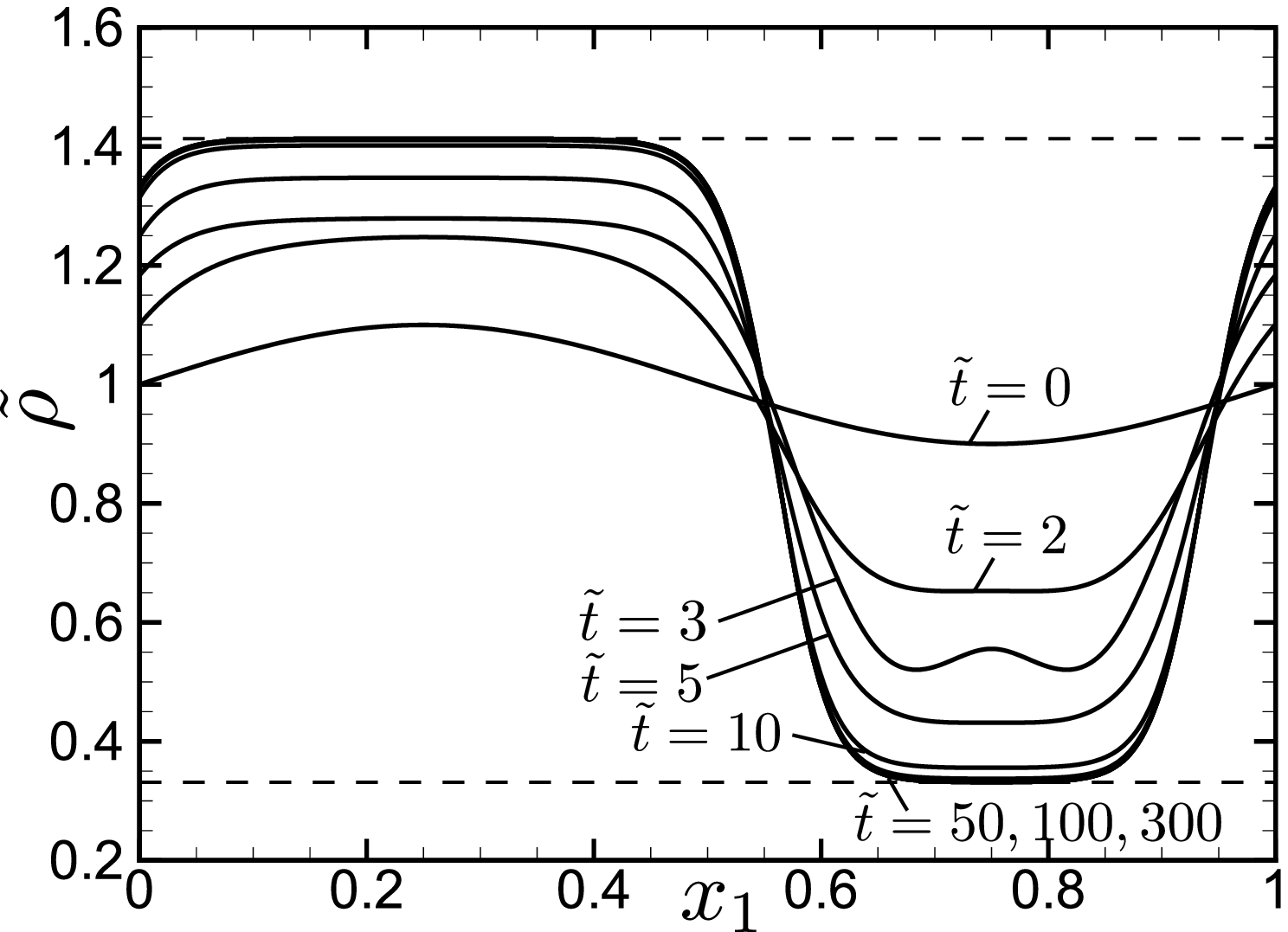}}

\caption{Time evolution of the density profile for cases L
and R for $(\sqrt{\pi}/2)\mathrm{Kn}=1$
and $\tilde{\kappa}/\tilde{b}=5\times10^{-4}$. (a) case L and (b)
case R. Top and bottom dashed lines in each panel indicate $\tilde{\rho}_{H}$
and $\tilde{\rho}_{L}$ predicted by the equiarea rule in Fig.~\ref{fig:diagram}(b).
See also Fig.~\ref{fig:Time-evolution-of}(b) (case C) for comparison.
\label{fig:Time-evolution-of-1-1}}
\end{figure}

Numerical computations have been performed mainly for cases C, L,
and R indicated in Fig.~\ref{fig:diagram}(a) by setting $\tilde{A}(\tilde{\rho})=1$.
Figure~\ref{fig:Time-evolution-of} shows the time evolution of the
density profile for case C for $(\sqrt{\pi}/2)\mathrm{Kn}=0.1$, $1$,
and $10$ and for case B2 for $(\sqrt{\pi}/2)\mathrm{Kn}=1$, where
$\tilde{\kappa}/\tilde{b}=5\times10^{-4}$ is common among the cases.
Roughly speaking, the profile in the evolution process was observed
to be a little more complicated at lower density region for smaller
$\tilde{\kappa}/\tilde{b}$ and for larger Knudsen number. Figure~\ref{fig:Time-evolution-of-1}
shows the evolution for $(\sqrt{\pi}/2)\mathrm{Kn}=1$ with different
values of $\tilde{\kappa}/\tilde{b}$. It is clear from the figure,
together with Fig.~\ref{fig:Time-evolution-of}(b), that the smaller
$\tilde{\kappa}/\tilde{b}$ is, the thinner the interface is. Figure~\ref{fig:Time-evolution-of-1-1}
shows the evolution in the cases L and R for $(\sqrt{\pi}/2)\mathrm{Kn}=1$.
It is seen that the dense region is thinner for case L, while it is
fatter for case R. Furthermore, the values of density plateaux after
a long time are different among the cases L, C, and R. Further comparisons
with Fig.~\ref{fig:Time-evolution-of} and Fig.~\ref{fig:nonplateaux}(b)
suggest that they are dependent on $\tilde{b}$ and $c$, neither
on $\mathrm{Kn}$ nor on $\tilde{\kappa}$. Indeed, the values of
plateaux, say $\tilde{\rho}_{L}$ and $\tilde{\rho}_{H}$, can be
predicted by the equiarea rule for the potential $\Phi$ (see Fig.~\ref{fig:diagram}(b)
and Ref.~\cite{TN17}), which is dependent only on $\tilde{b}$ and
$c$.

\begin{figure*}
\centering
\subfigure[]{\includegraphics[clip,scale=0.4]{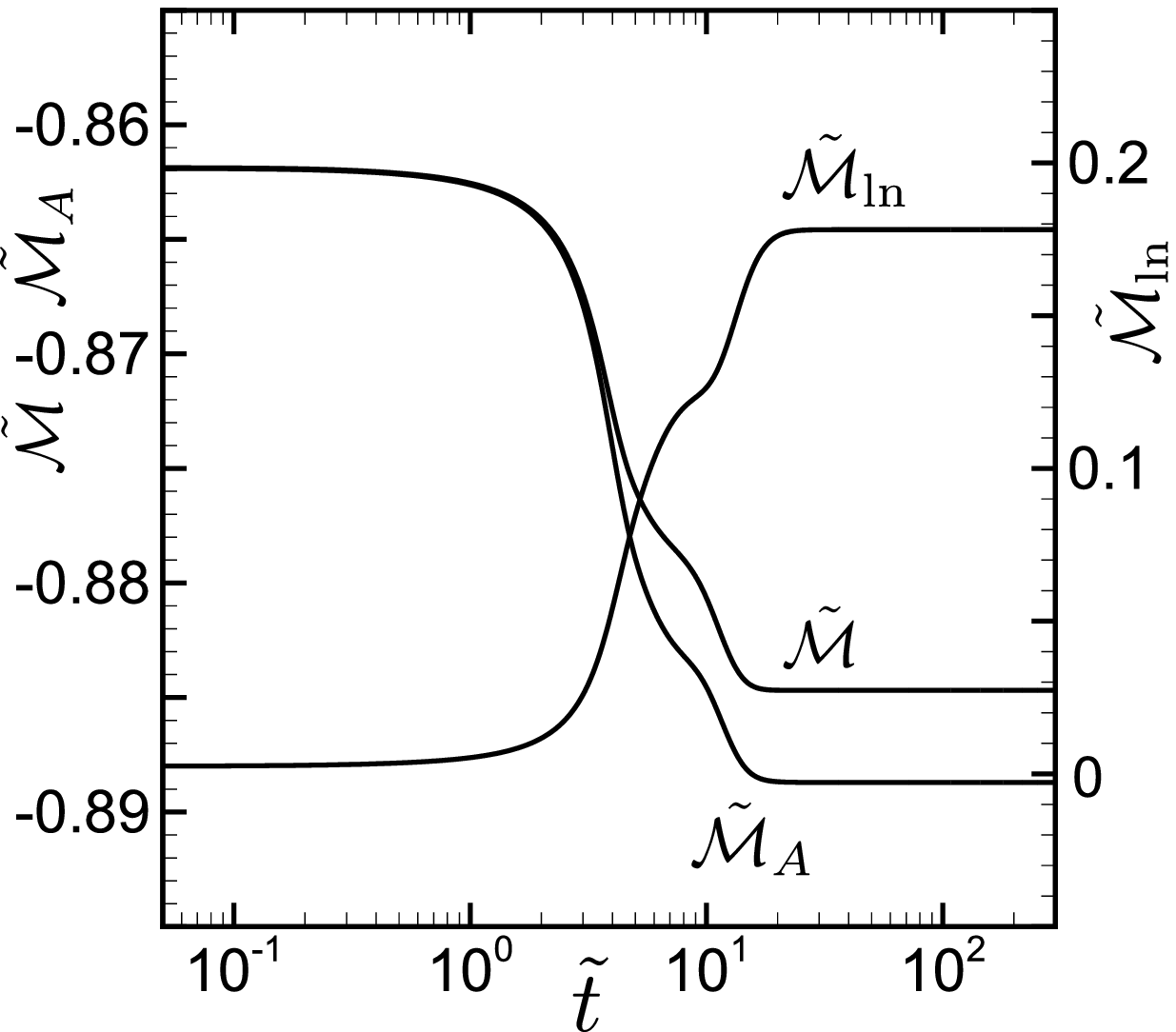}}\quad
\subfigure[]{\includegraphics[bb=0bp 0bp 349bp 312bp,clip,scale=0.4]{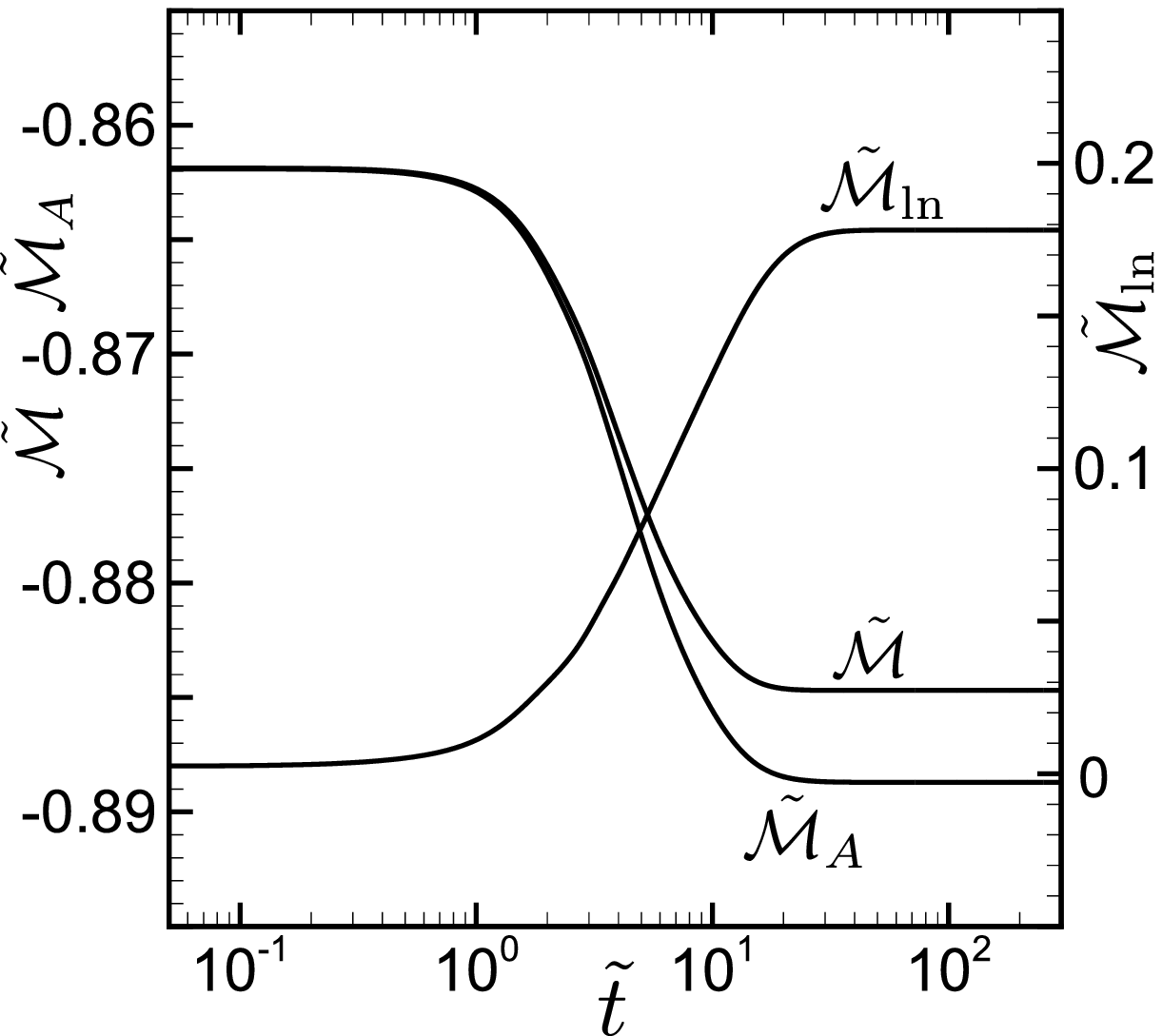}}\quad
\subfigure[]{\includegraphics[bb=0bp 0bp 359bp 306bp,clip,scale=0.4]{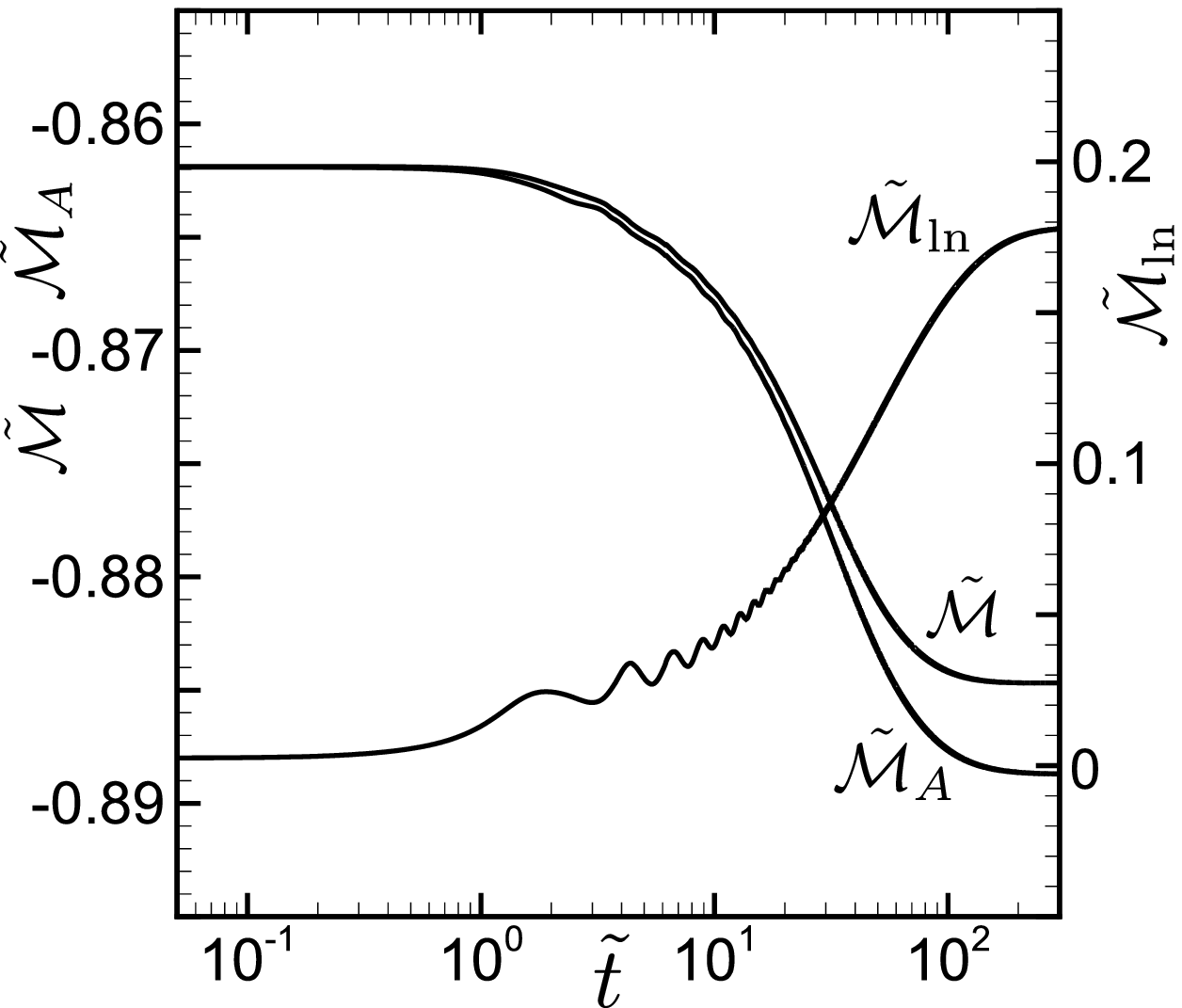}}\\
\subfigure[]{ \includegraphics[bb=0bp 0bp 354bp 309bp,clip,scale=0.4]{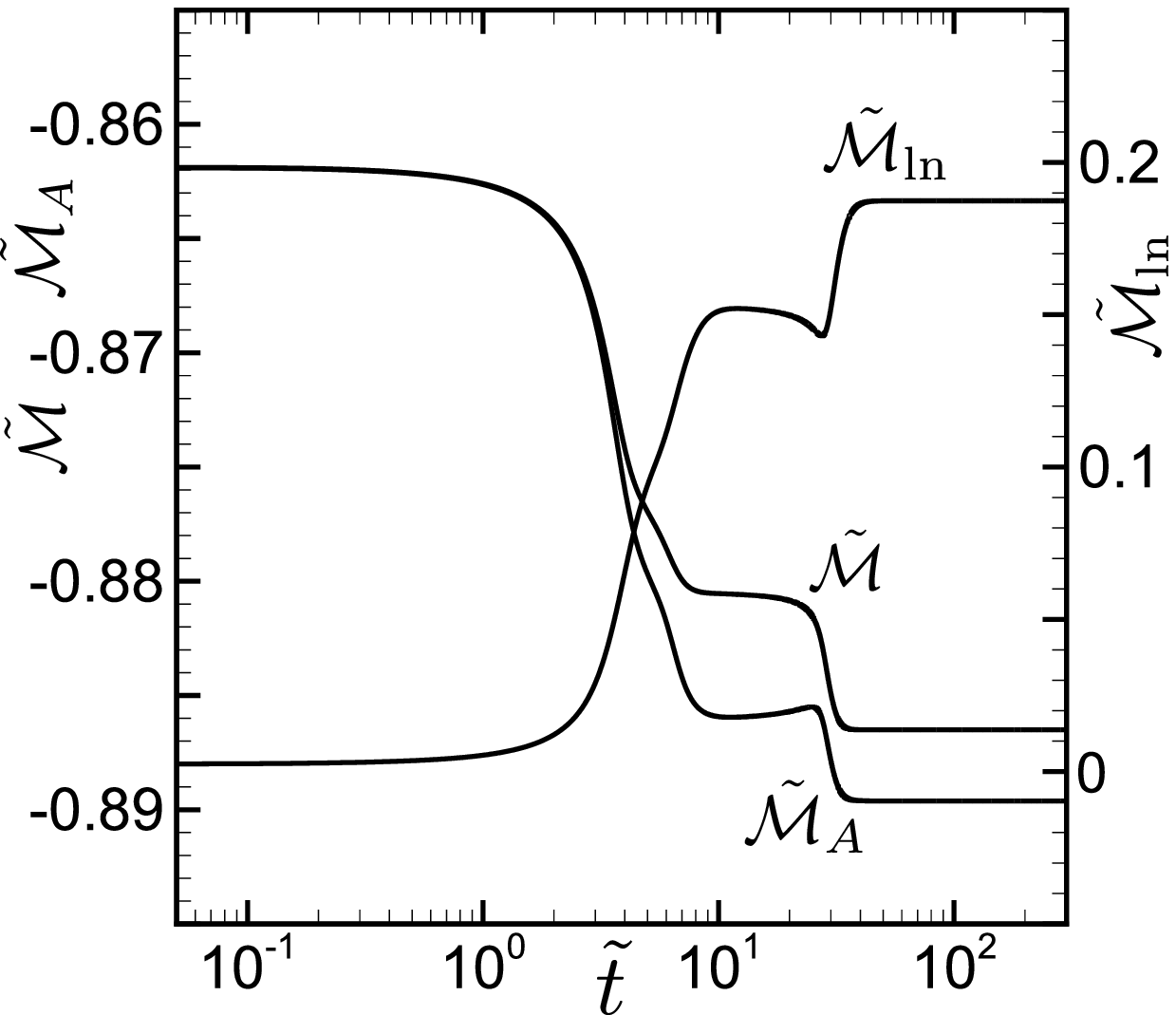}}\quad
\subfigure[]{\includegraphics[bb=0bp 0bp 384bp 319bp,clip,scale=0.4]{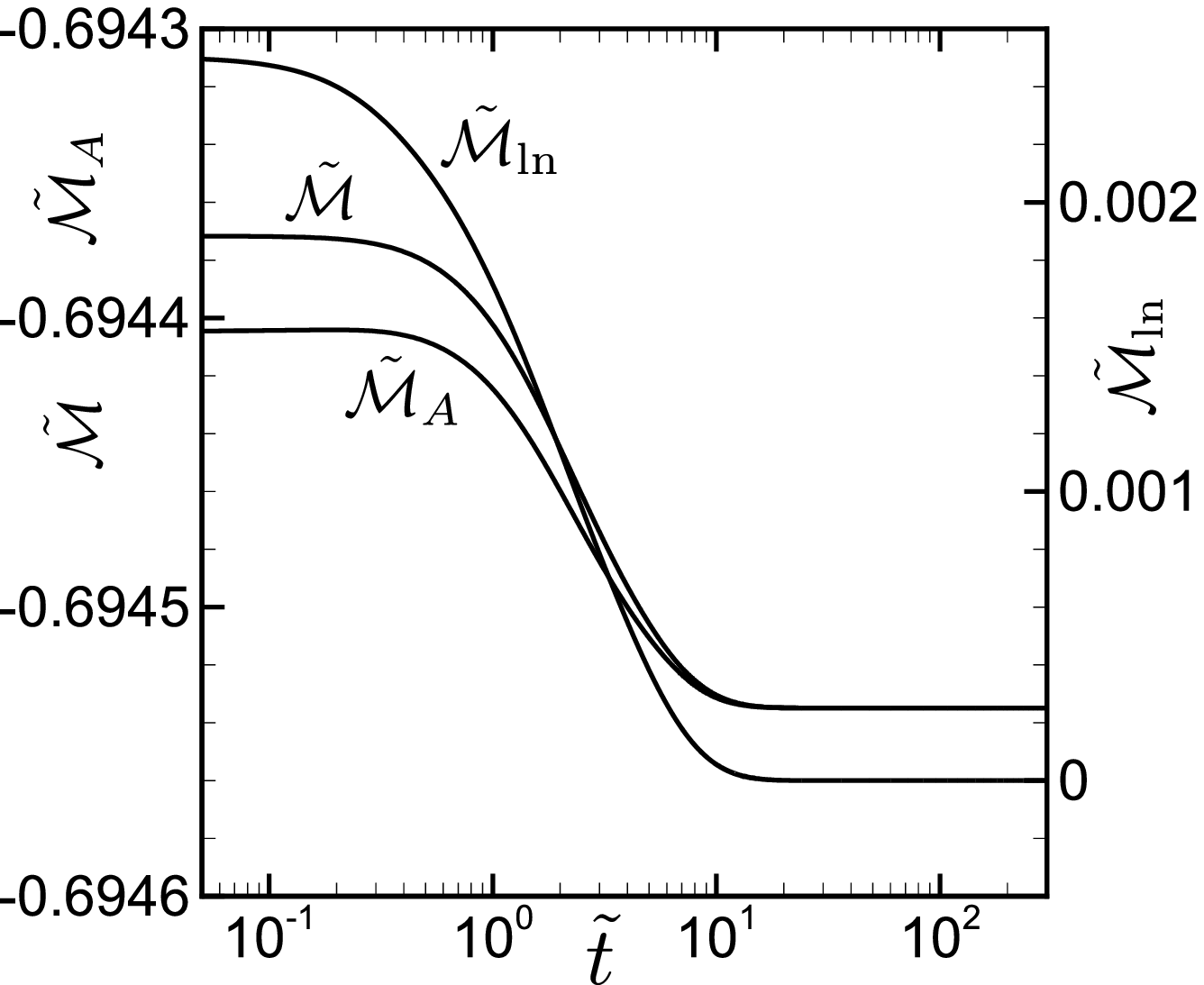}}\quad
\subfigure[]{\includegraphics[bb=0bp 0bp 354bp 309bp,clip,scale=0.4]{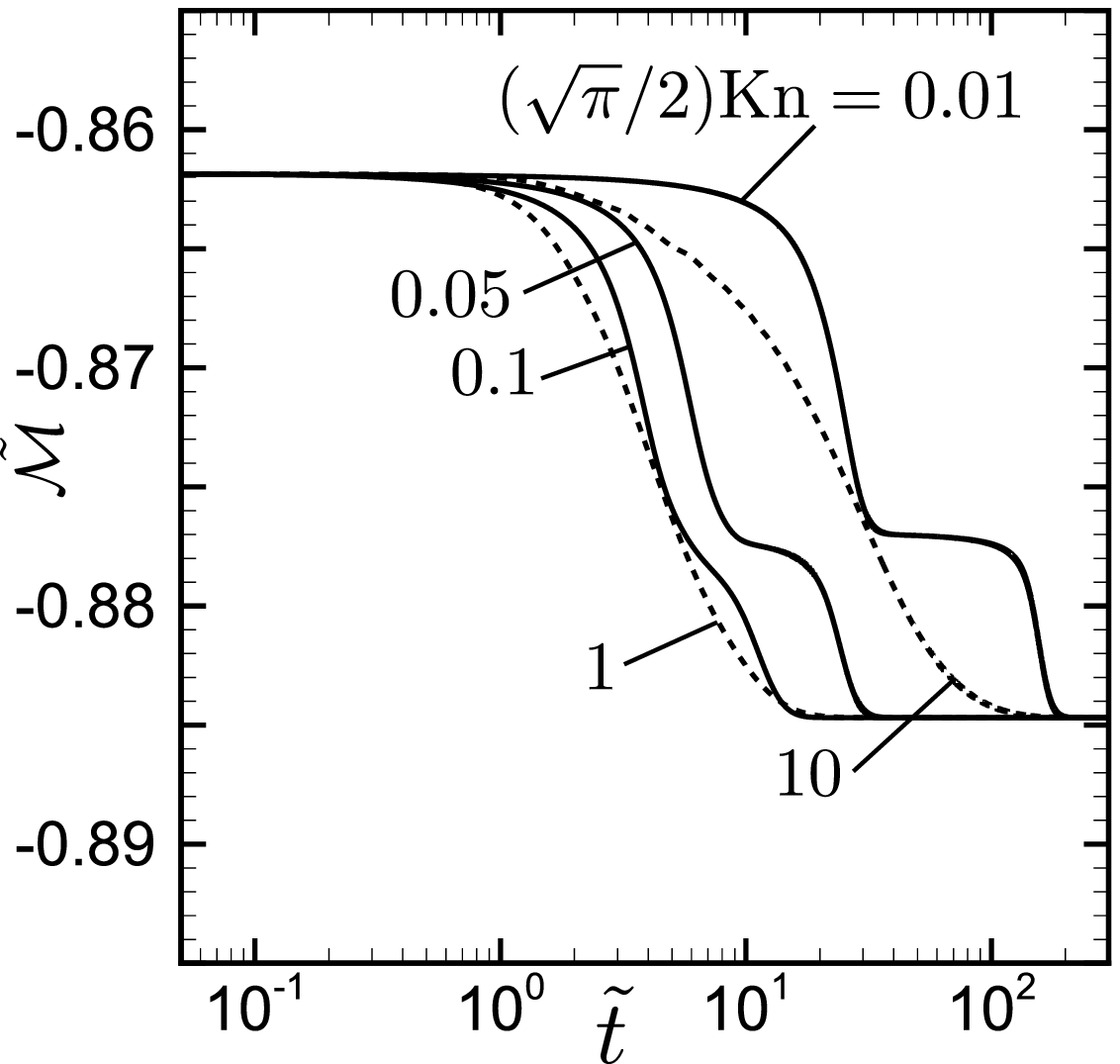}}

\caption{Time evolution of $\tilde{\mathcal{M}}$, $\tilde{\mathcal{M}}_{A}$,
and $\tilde{\mathcal{M}}_{\mathrm{ln}}$. Panels (a)\textendash (c)
are case C with $\tilde{\kappa}/\tilde{b}=5\times10^{-4}$ for different
Knudsen numbers: (a) $(\sqrt{\pi}/2)\mathrm{Kn}=0.1$, (b) $(\sqrt{\pi}/2)\mathrm{Kn}=1$,
and (c) $(\sqrt{\pi}/2)\mathrm{Kn}=10$. Panel (d) is case C with
$\tilde{\kappa}/\tilde{b}=3\times10^{-4}$ and $(\sqrt{\pi}/2)\mathrm{Kn}=0.1$,
while Panel (e) is case B2 with $\tilde{\kappa}/\tilde{b}=5\times10^{-4}$
and $(\sqrt{\pi}/2)\mathrm{Kn}=1$.
Panel (f) is case C with $\tilde{\kappa}/\tilde{b}=5\times10^{-4}$
for $(\sqrt{\pi}/2)\mathrm{Kn}=0.01, 0.05, 0.1, 1$ and $10$. \label{fig:minimization}}
\end{figure*}

\begin{figure*}
\centering
\subfigure[]{\includegraphics[bb=0bp -3bp 429bp 309bp,clip,scale=0.37]{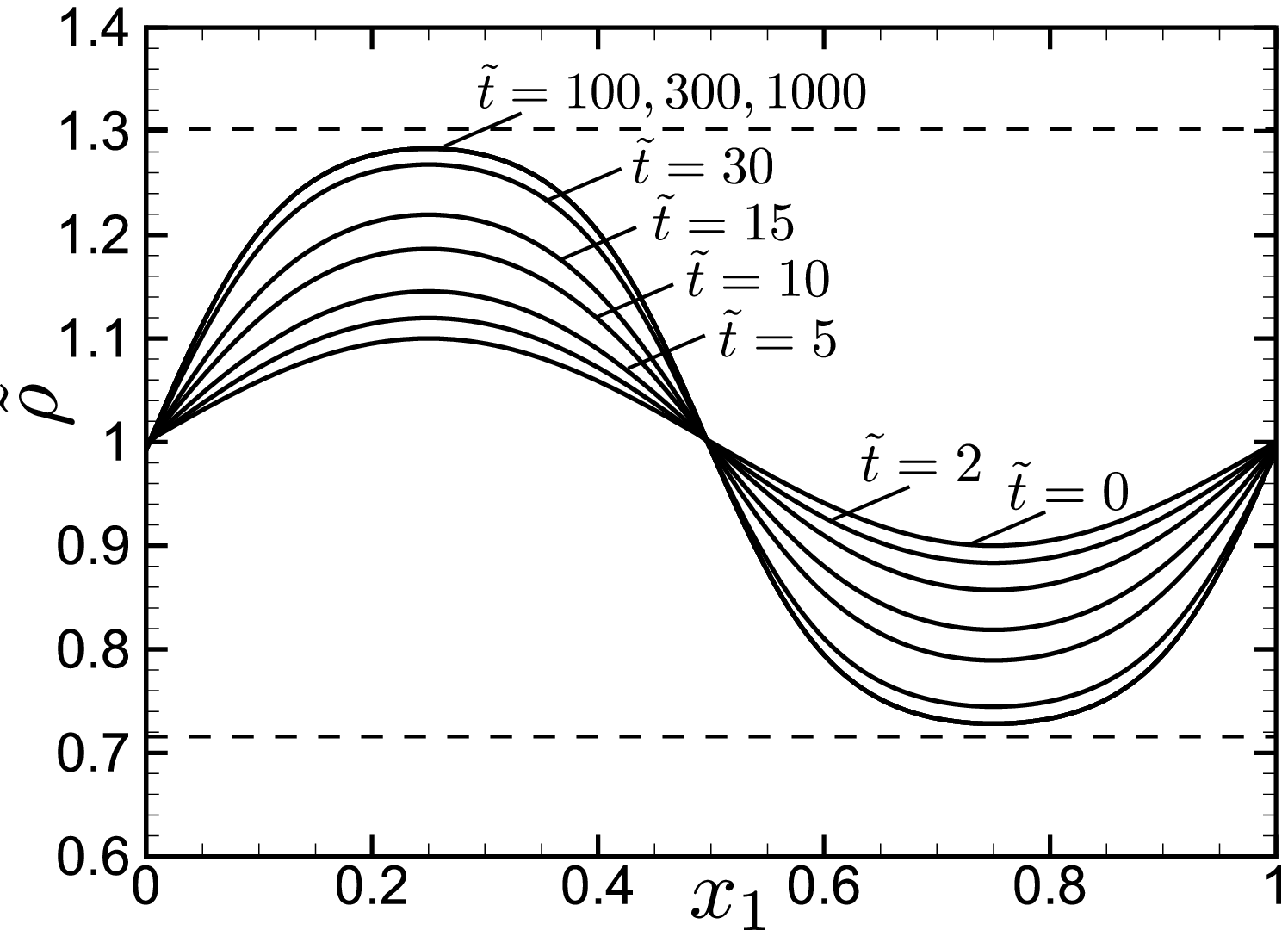}}\
\subfigure[]{\includegraphics[bb=0bp -3bp 443bp 308bp,clip,scale=0.37]{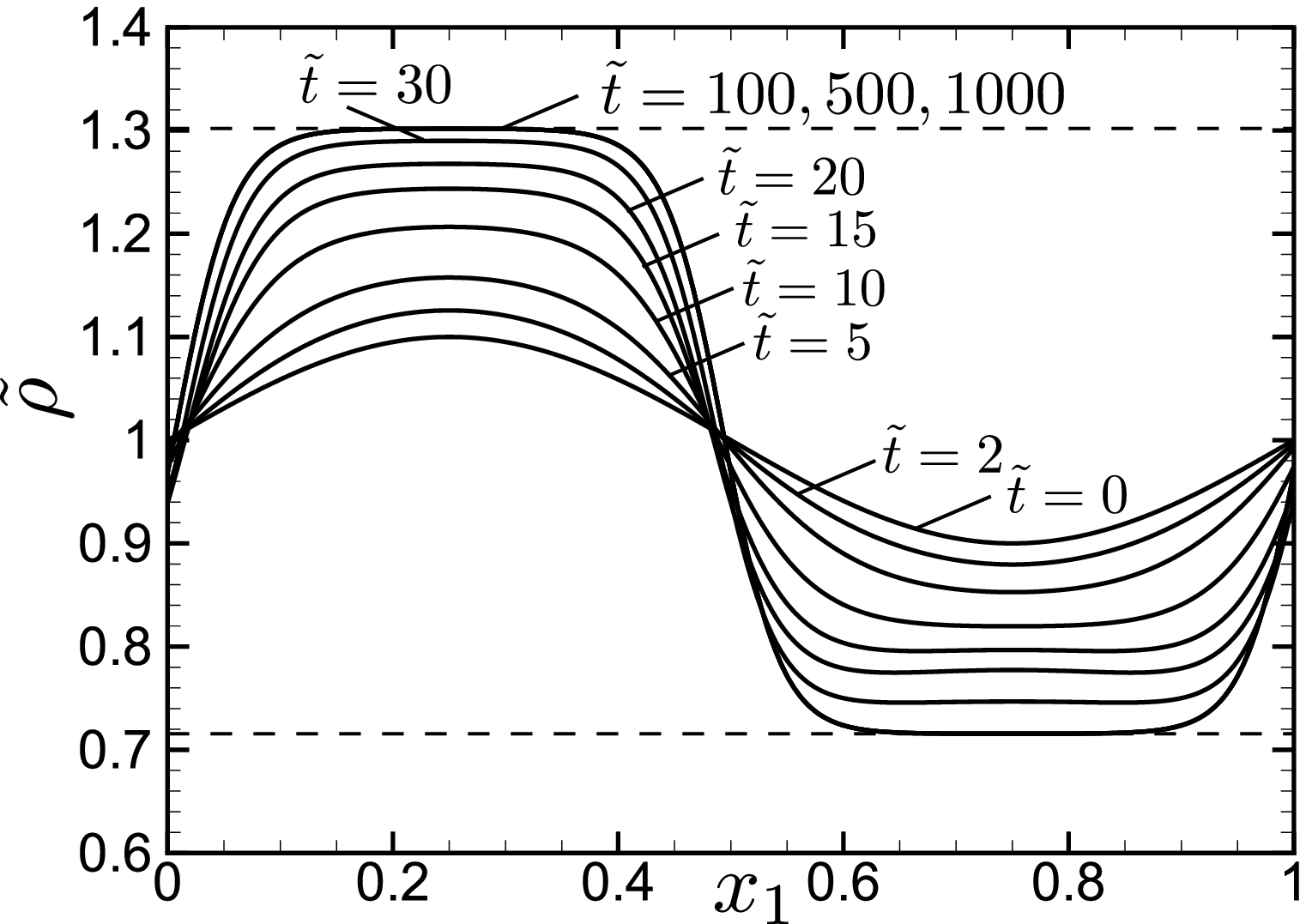}}\
\subfigure[]{\includegraphics[bb=0bp 0bp 353bp 314bp,clip,scale=0.37]{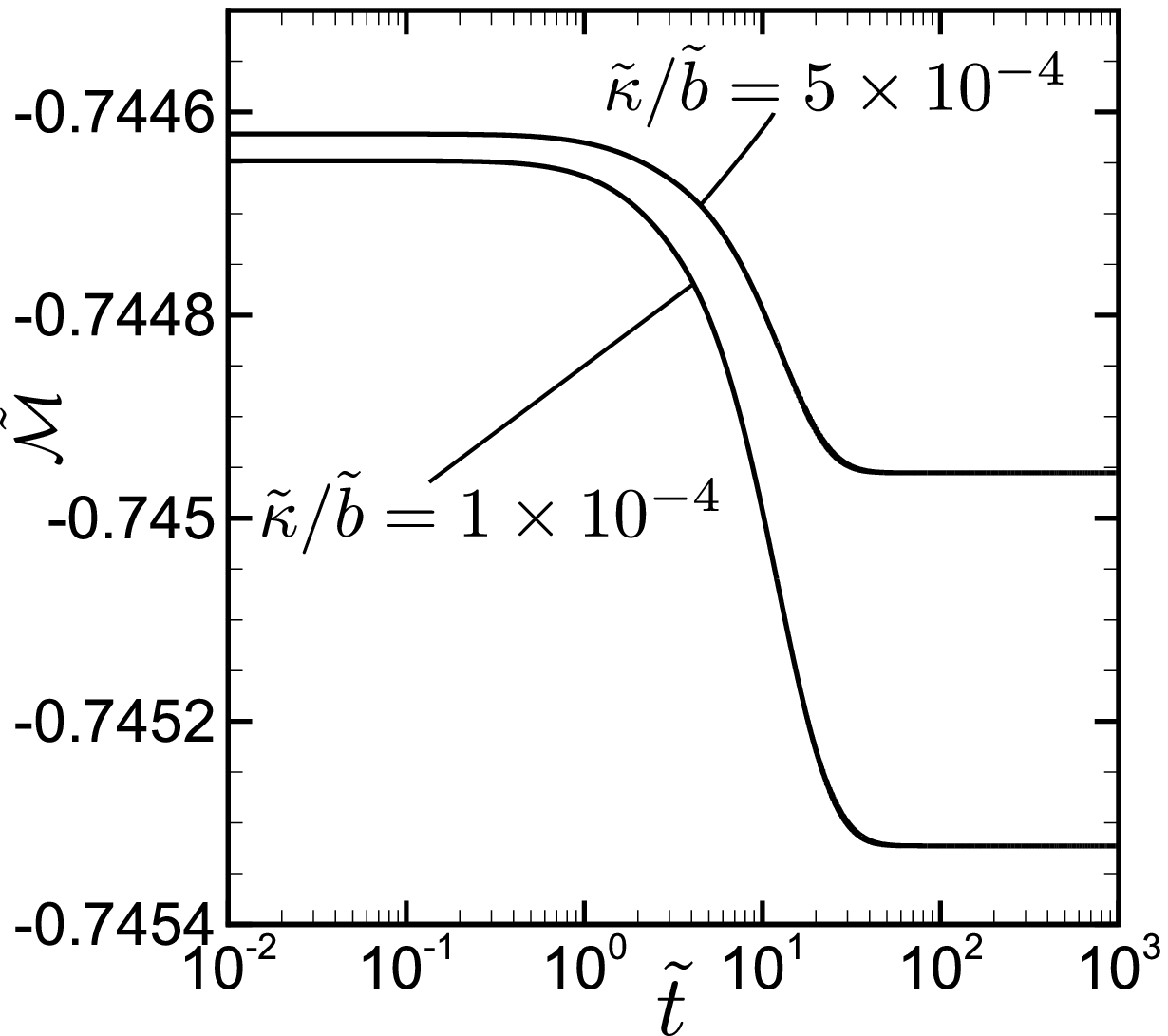}}
\caption{Time evolution of the density and functional $\tilde{\mathcal{M}}$ in case
A. (a) $\tilde{\rho}$ for $\tilde{\kappa}/\tilde{b}=5\times10^{-4}$,
(b) $\tilde{\rho}$ for $\tilde{\kappa}/\tilde{b}=1\times10^{-4}$,
and (c) $\tilde{\mathcal{M}}$ for both values of $\tilde{\kappa}/\tilde{b}$,
where $\mathrm{Kn}$ is commonly set as $(\sqrt{\pi}/2)\mathrm{Kn}=1$.
Top and bottom dashed lines in (a) and (b) indicate $\tilde{\rho}_{H}$
and $\tilde{\rho}_{L}$, respectively. \label{fig:nonplateaux}}
\end{figure*}

Time evolutions of the functional $\mathcal{\tilde{M}}$ are shown
in Fig.~\ref{fig:minimization} for various parameters. Here, the
integration in space occurring in $\mathcal{\tilde{M}}$ should be
understood as $\int_{\tilde{D}}\cdots d\bm{x}=\int_{0}^{1}\cdots dx_{1}$.
As expected, $\tilde{\mathcal{M}}$ is always monotonic and decreasing
in time, while individual parts of it, say
\begin{align*}
\tilde{\mathcal{M}}_{\mathrm{ln}} & \equiv\int_{\tilde{D}}\int\tilde{f}\ln\frac{\tilde{f}}{E}d\bm{\zeta}d\bm{x},\displaybreak[0]\\
\tilde{\mathcal{M}}_{A} & \equiv\tilde{\mathcal{M}}_{\mathrm{ln}}+2\int_{\tilde{D}}\int\tilde{\Phi}_{S}d\tilde{\rho}d\bm{x},\displaybreak[0]\\
\tilde{\mathcal{M}}_{L} & \equiv\int_{\tilde{D}}\tilde{\rho}\tilde{\Phi}_{L}d\bm{x}=\tilde{\kappa}\int_{\tilde{D}}\Big(\frac{\partial\tilde{\rho}}{\partial x_{i}}\Big)^{2}d\bm{x}(=\tilde{\mathcal{M}}-\tilde{\mathcal{M}}_{A}),
\end{align*}
are not necessarily monotonic. In viewing Figs.~\ref{fig:minimization}(a)\textendash (c),
it is likely that the transition takes more time for larger Knudsen number.
In the meantime,
the proper scaling that leads to the Cahn--Hilliard type equation in the continuum limit \cite{TN17}
is to set the Strouhal number $\mathrm{Sh}$ as the order of the Knudsen number $\mathrm{Kn}$.
This implies that the transition will take more time as $\mathrm{Kn}$ becomes smaller,
which seems to conflict with the above numerical evidence.
Actually, however, there is no confliction.
The results of additional computations
for $(\sqrt{\pi}/2)\mathrm{Kn}=0.05$ and $0.01$ show
that the transition indeed turns to take more time as $\mathrm{Kn}$ becomes much smaller,
as is shown in Fig.~\ref{fig:minimization}(f).

Going back to the density profiles, and let us now discuss how the
values of density plateaux after a long time are determined. Since
$\tilde{\mathcal{M}}$ is monotonically decreasing, the system approaches
to its stationary state, which motivates us to consider the variational
problem of $\tilde{\mathcal{M}}$ with respect to $\tilde{f}$. Variational
problem of $\tilde{\mathcal{M}}$ with respect to $\tilde{f}$ followed
by van Kampen's interpretation \citep{vK64} leads to the equiarea
rule in Fig.~\ref{fig:diagram}(b) for determining those values in
one-dimensional case. Let us consider the first variation of (\ref{eq:Mdless})
under the constraint of $\int_{\tilde{D}}\tilde{\rho}d\bm{x}\equiv\int_{\tilde{D}}\int\tilde{f}d\bm{\zeta}d\bm{x}=V_{\tilde{D}}$:
\begin{align*}
 & \delta\tilde{\mathcal{M}}\\
= & \int_{\tilde{D}}\int\{1+\ln\frac{\tilde{f}}{E}+2\tilde{\Phi}_{S}(\tilde{\rho})-\tilde{\kappa}\frac{\partial^{2}\tilde{\rho}}{\partial x_{i}^{2}}-\tilde{\rho}\tilde{\kappa}\frac{\partial^{2}}{\partial x_{i}^{2}}-\lambda\}\delta\tilde{f}d\bm{\zeta}d\bm{x}\displaybreak[0]\\
= & \int_{\tilde{D}}\int\{1+\ln\frac{\tilde{f}}{E}+2\tilde{\Phi}_{S}(\tilde{\rho})-2\tilde{\kappa}\frac{\partial^{2}\tilde{\rho}}{\partial x_{i}^{2}}-\lambda\}\delta\tilde{f}d\bm{\zeta}d\bm{x},
\end{align*}
where $\lambda$ is the Lagrange multiplier, which is constant in
$\bm{x}$ and $\bm{\zeta}$. Thus, it holds that
\[
1+\ln\frac{\tilde{f}}{E}+2\tilde{\Phi}_{S}(\tilde{\rho})-2\tilde{\kappa}\frac{\partial^{2}\tilde{\rho}}{\partial x_{i}^{2}}-\lambda=0,
\]
at the stationary state. Then, taking into account the balance among
the component terms, $\ln(\tilde{f}/E)$ is found to be independent
of $\bm{\zeta}$, leading to $\tilde{f}=\tilde{\rho}E$. Hence, the
above condition is reduced to
\[
1+\ln\tilde{\rho}+2\tilde{\Phi}_{S}(\tilde{\rho})-2\tilde{\kappa}\frac{\partial^{2}\tilde{\rho}}{\partial x_{i}^{2}}-\lambda=0,
\]
or equivalently
\begin{align*}
 & \tilde{\kappa}\frac{\partial^{2}\tilde{\rho}}{\partial x_{i}^{2}}=\Phi(\tilde{\rho})-\frac{1}{2}\lambda,\displaybreak[0]\\
 & \Phi(\tilde{\rho})\equiv\frac{1}{2}+\frac{1}{2}\ln\tilde{\rho}+\tilde{\Phi}_{S}=-\tilde{a}\tilde{\rho}+\frac{1}{2}\frac{1}{1-\tilde{b}\tilde{\rho}}+\frac{1}{2}\ln\frac{\tilde{\rho}}{1-\tilde{b}\tilde{\rho}}.
\end{align*}
By interpreting this condition as a motion of point mass following
van Kampen (see Refs.~\cite{vK64,TN17}), we can conclude that the
equiarea rule
\begin{subequations}\label{eq:equiarea}
\begin{align}
\int_{\tilde{\rho}_{L}}^{\tilde{\rho}_{H}}&\Phi(r)dr=\frac{\lambda}{2}(\tilde{\rho}_{H}-\tilde{\rho}_{L}),
\intertext{with }
&\Phi(\tilde{\rho}_{H})=\Phi(\tilde{\rho}_{L})=\frac{\lambda}{2},
\end{align}
\end{subequations}
applies, where $\tilde{\rho}_{H}$ and $\tilde{\rho}{}_{L}$ respectively
denote larger and smaller values of density plateaux. This rule determines
$\tilde{\rho}_{H}$ and $\tilde{\rho}{}_{L}$ as sketched in Fig.~\ref{fig:diagram}(b).
Note that the curve of $\Phi$ is dependent only on $\tilde{b}$;
once $\tilde{a}$ (or $c$) is fixed, so are the values of $\tilde{\rho}_{H}$
and $\tilde{\rho}_{L}$.

\begin{figure*}
  \centering
  \subfigure[]{\includegraphics[bb=0bp 0bp 429bp 312bp,clip,scale=0.37]{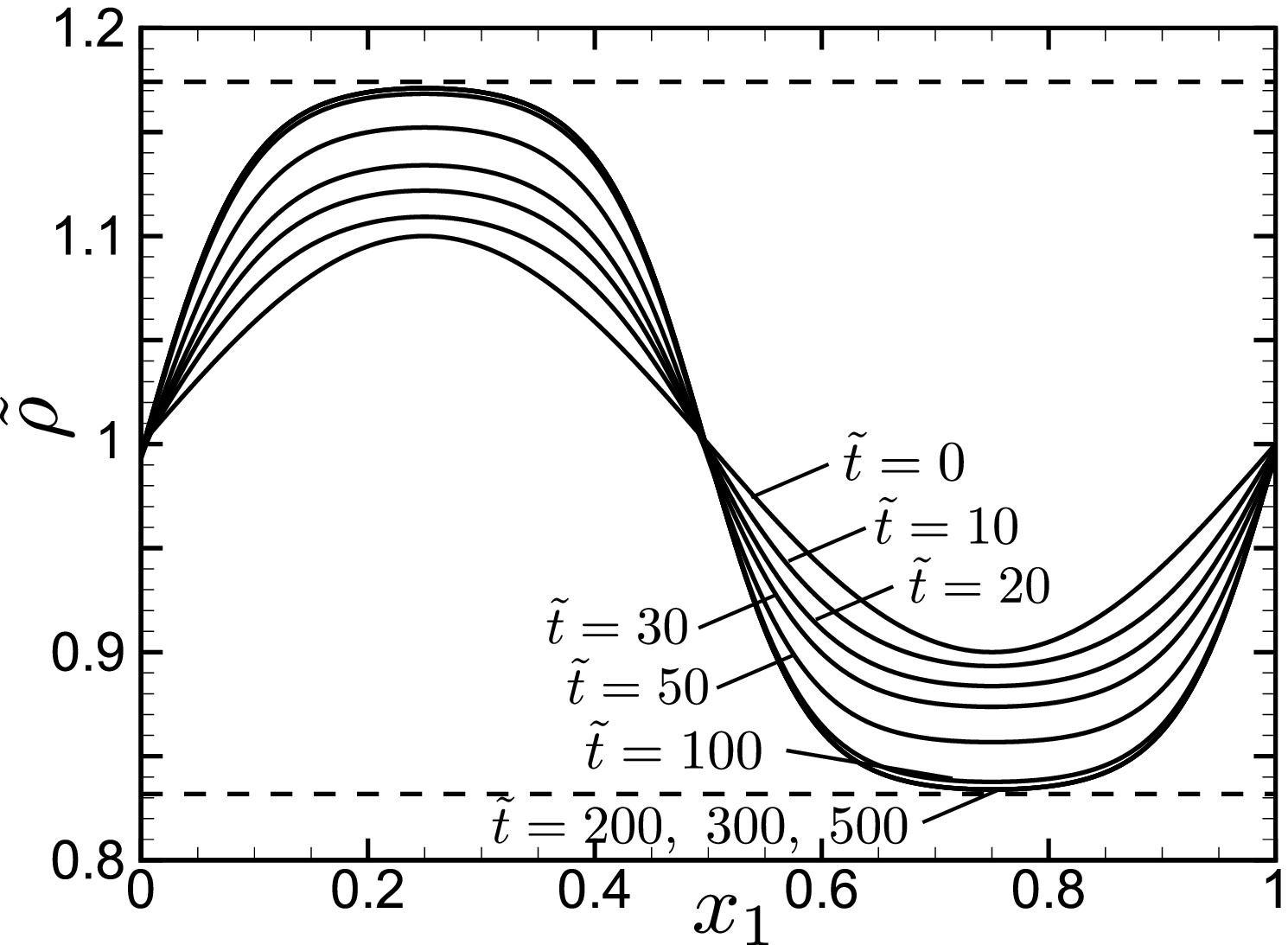}}\quad
  \subfigure[]{\includegraphics[bb=0bp 0bp 443bp 308bp,clip,scale=0.37]{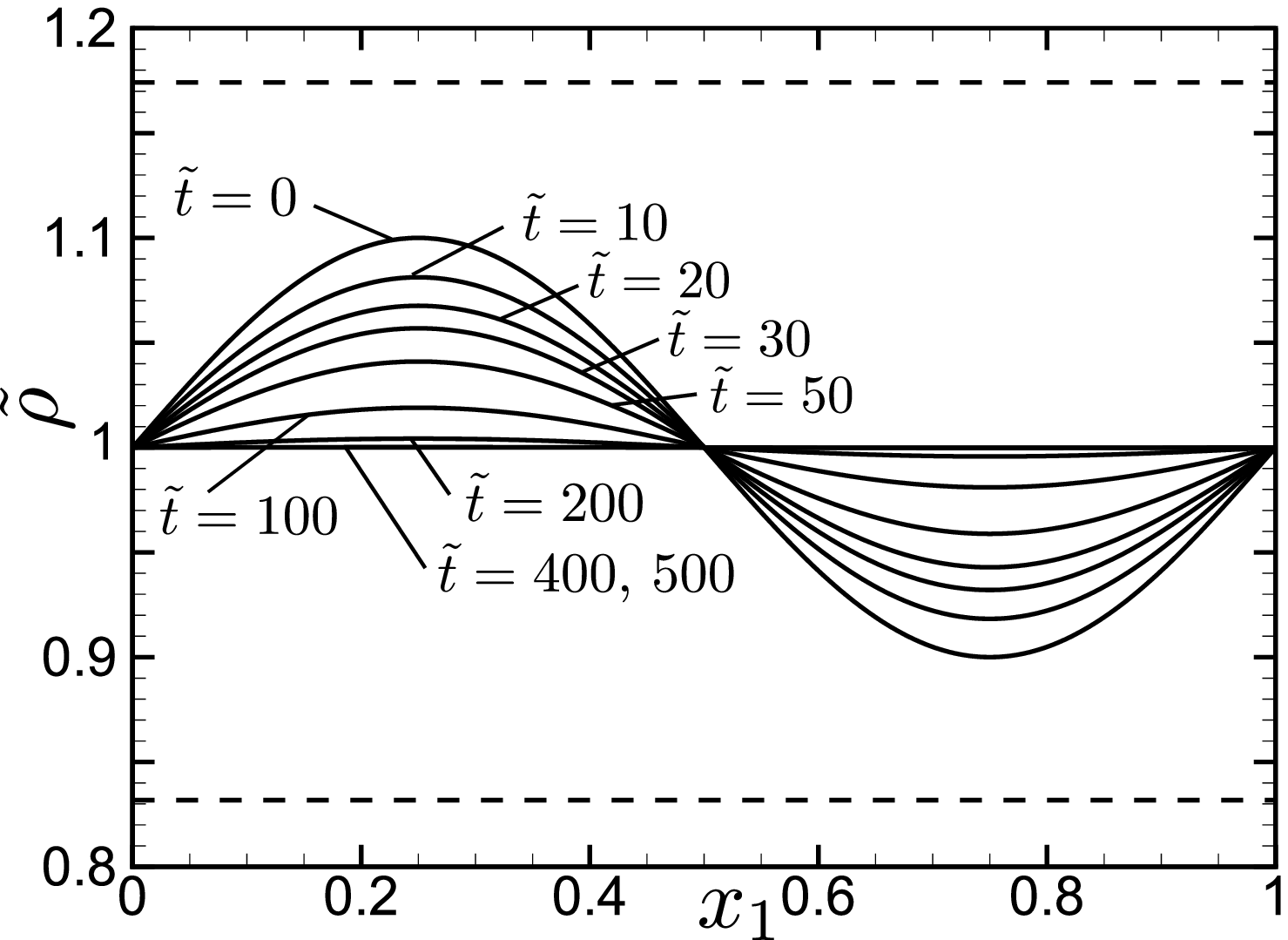}}\\
  \subfigure[]{\includegraphics[bb=0bp 0bp 823bp 310bp,clip,scale=0.42]{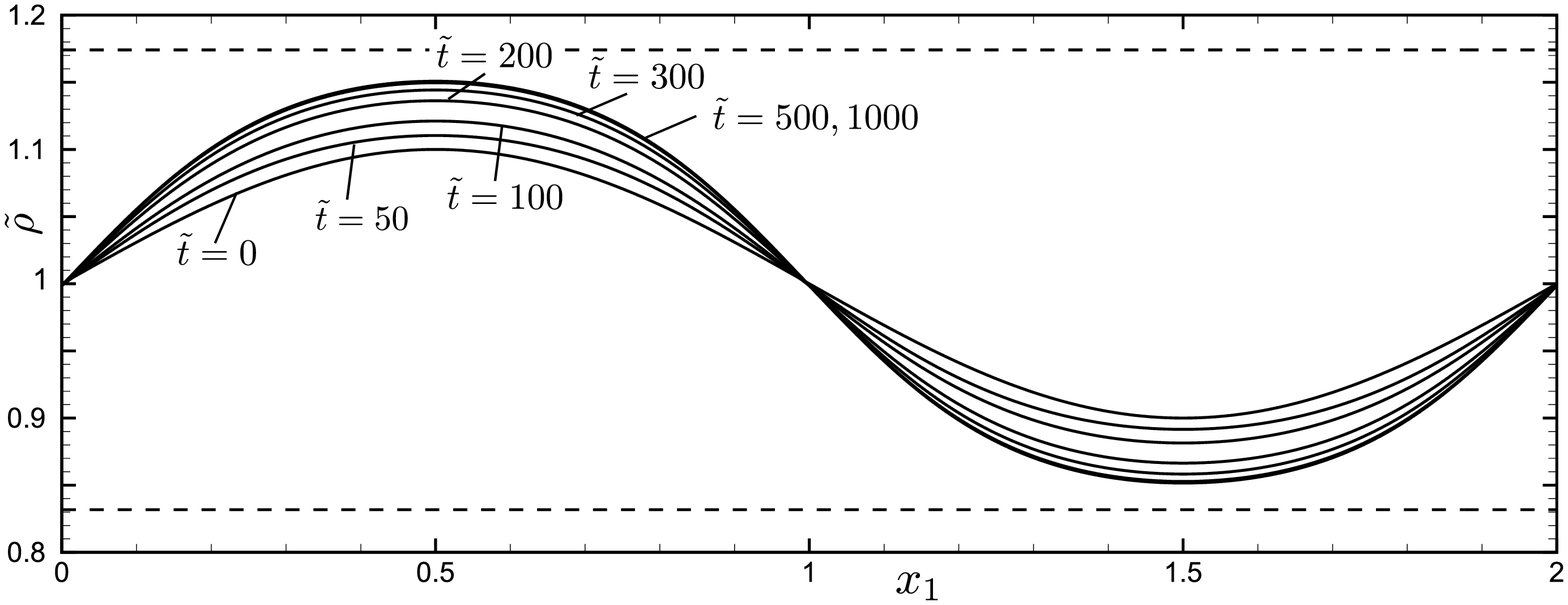}}
  \caption{Influence of the domain periodicity:
  time evolution of the density profile for case D
  for $(\sqrt{\pi}/2)\mathrm{Kn}=0.1$.
  (a) $\tilde{\kappa}/\tilde{b}=1\times10^{-4}$ with period $1$,
  (b) $\tilde{\kappa}/\tilde{b}=1\times10^{-3}$ with period $1$,
  (c) $\tilde{\kappa}/\tilde{b}=1\times10^{-3}$ with period $2$.
  In (c), the sinusoidal initial perturbation has a period $2$.
  Top and bottom dashed lines in each panel indicate $\tilde{\rho}_{H}$
  and $\tilde{\rho}_{L}$, respectively. \label{fig7}}
\end{figure*}

Finally,
we shall remark a couple of unexpected phenomena
that were found numerically in the above neighborhood of the neutral curve.
Firstly, a clear separation
of plateaux and interfaces might not be achieved in the above neighborhood
of the neutral curve, if $\tilde{\kappa}/\tilde{b}$ is not sufficiently
small. Figure~\ref{fig:nonplateaux}(a) demonstrates such an example
(case A, $\tilde{\kappa}/\tilde{b}=5\times10^{-4}$ and $(\sqrt{\pi}/2)\mathrm{Kn}=1$).
In the case, the phase transition is incomplete and the evolution
ceases with a smooth profile as in Fig.~\ref{fig:nonplateaux}(a).
In the meantime, the interface ought to be thinner for smaller $\tilde{\kappa}$
and the transition is more likely to be completed. Indeed, for case
A with $\tilde{\kappa}/\tilde{b}=1\times10^{-4}$ and $(\sqrt{\pi}/2)\mathrm{Kn}=1$,
a clear separation of different plateaux and interfaces is observed,
as in Fig.~\ref{fig:nonplateaux}(b). Then, the equiarea rule revives
for the prediction of the values of density plateaux.
Secondly, a uniform equilibrium state can be stable in the above neighborhood of the neutral curve
as shown in Fig.~\ref{fig7}(b), which would be more striking.
This is, however, understandable, if we take into account the domain periodicity
in the linear stability analysis in Sec.~\ref{sec:Linear-stability-of}.
As we have already mentioned it at the end of Sec.~\ref{sec:Linear-stability-of},
the periodicity of space domain has not been taken into account
in the derivation of the unstable condition \eqref{eq:instability condition}.
In order to take it into account,
we need to go back to \eqref{eq:cond_a}
and to use that the admitted modes should meet the condition $\lambda >2\pi$.
Then, the unstable condition is found to take the following modified form
\begin{equation}\label{md_unst}
  c>\frac{1}{2\tilde{b}(1-\tilde{b})^2}+\frac{\tilde\kappa}{\tilde{b}}(2\pi)^2.
\end{equation}
It is readily seen that all the cases for which the transition,
including the \textit{incomplete} one, is observed
meet the above condition, while all the cases
for which the uniform equilibrium state is observed stable do not.
For instance, case D with $(\sqrt{\pi}/2)\mathrm{Kn}=0.1$ shown in Fig.~\ref{fig7}(a)
is in modified linear unstable region when $\tilde{\kappa}/\tilde{b}=1\times 10^{-4}$.
It is, however, shifted indeed into the modified linear stable region
by changing the value of $\tilde{\kappa}/\tilde{b}$ to $1\times 10^{-3}$.
Then, the uniform equilibrium state becomes stable,
as is demonstrated numerically in Fig.~\ref{fig7}(b).
By the same reason, in case D with $\tilde{\kappa}/\tilde{b}=1\times 10^{-3}$,
the uniform equilibrium state becomes unstable, if the domain period is doubled.
It is because the upperbound of admitted mode period is accordingly doubled;
consequently the unstable condition becomes
\begin{equation*}
  c>\frac{1}{2\tilde{b}(1-\tilde{b})^2}+\frac{\tilde\kappa}{\tilde{b}}\pi^2,
\end{equation*}
and case D falls into a new unstable region.
It is demonstrated numerically in Fig.~\ref{fig7}(c).


\section{Conclusion\label{sec:Conclusion}}

In the present paper, we have investigated the stability of uniform
equilibrium states in the kinetic regime. The linear stability analysis
shows that the linear unstable condition is the same as that of the
Cahn\textendash Hilliard type equation in Ref.~\cite{TN17}, irrespective
of the Knudsen number. The condition is not affected by $\kappa$
(or $\tilde{\kappa}$) as well, the measure of the collective long-range
interaction effect occurring in $\Phi_{L}$ (or $\tilde{\Phi}_{L}$).
By numerical computations we have indeed observed the phase transition
after perturbing uniform equilibrium states that lie above the neutral
curve in the diagram of Fig.~\ref{fig:diagram}(a). We did not observe
unstable uniform equilibrium states below this curve (cases B1, B2,
and B3 in Fig.~\ref{fig:diagram}(a)), though the computed cases
are limited. The numerical results show that $\kappa$ (or $\tilde{\kappa}$)
mainly affects the thickness of the interface between different phases;
the smaller $\kappa$ (or $\tilde{\kappa}$) is, the thinner the interface
is. The Knudsen number affects the transition process but does not
affect the values of density plateaux at the final states. They are
determined by another diagram, i.e., the equiarea rule, in Fig.~\ref{fig:diagram}(b),
as far as the interface is thin enough for the clear separation of
different plateaux to emerge. The numerical results further show that
the functional $\mathcal{M}$ (or $\tilde{\mathcal{M}}$) indeed monotonically
decreases as predicted in Ref.~\cite{TN17}, though its component
terms are not necessarily monotonic in the transition process.

\paragraph*{Acknowledgements}

The present work is supported in part by JSPS KAKENHI Grant Number
17K18840.

\appendix
\section{On the form of $\Phi_L$ and $\Phi_S$\label{APP:0}}

Here we repeat the explanation about the specific form of $\Phi_L$ and $\Phi_S$
briefly for the sake of convenience of the reader,
though it has already been given in Ref.~\cite{TN17}.

The self-consistent force potential $\phi$,
which is of the conventional Vlasov-type, is split into attractive
and repulsive parts. The attractive part, $\Phi_{A}$, is of long-range,
while the repulsive part, $\Phi_{R}$, is of short-range and is a
function of the local density $\rho$. By the latter and a part of
the former, we intend to reproduce a non-ideal gas feature under the
isothermal approximation, which is represented by the potential $\Phi_{S}$.
Excluding effect by the repulsive force is usually included in the
collision term with detailed collision dynamics, like in the Enskog
equation. Hence, the simplification by combining
the mean-field repulsive potential and the simplified role of the
collision term is the main difference from the existing model \cite{G71,FGL05}.

The attractive mean field is expressed by
\begin{align}
 m\Phi_{A}(t,\bm{X})
=&\int_{\mathbb{R}^3}\Psi(|\bm{r}|)\{\rho(t,\bm{X}+\bm{r})-\rho(t,\bm{X})\}d\bm{r}\nonumber \\
& +\int_{\mathbb{R}^3}\Psi(|\bm{r}|)d\bm{r}\rho(t,\bm{X})\nonumber \\
\equiv & m\Phi_{L}[\rho]+\int_{\mathbb{R}^3}\Psi(|\bm{r}|)d\bm{r} \rho(t,\bm{X}),
\label{eq:longrange}
\end{align}
where $m\Psi$ is the attractive intermolecular potential and is assumed
to be isotropic. Here, $\Phi_{L}$ may be considered as a contribution
from the long tail to the total attractive potential. The subtracted
part $\int_{\mathbb{R}^3}\Psi(|\bm{r}|)d\bm{r}\ \rho(t,\bm{X})$ will be combined
with the repulsive part $\Phi_R$ to form the residue $m\Phi_{S}$ in the total
self-consistent potential $m\phi$:
\begin{equation}
m\Phi_{S}=m\Phi_{R}+\{\int_{\mathbb{R}^3}\Psi(|\bm{r}|)d\bm{r}\}\rho(t,\bm{X}),
\end{equation}
the functional form of which is to be determined from the van
der Waals equation of state.
Since $\Phi_S$ is of short range, we are motivated to treat this as a local
(or internal) variable, thereby related to the stress tensor (and the static pressure)
through the momentum balance equation.
This is the key idea behind our phenomenological determination of $\Phi_S$ from the equation of state
 (See Sec.~3 of Ref.~\cite{TN17} for the specific construction).

When $\Psi$ decays fast in the system size as usually expected, the
variation of $\rho$ is moderate in that scale and the Taylor expansion
is allowed to yield
\begin{align}
  \Phi_{L}[\rho](t,\bm{X})
=&\frac{1}{m}\int_{\mathbb{R}^3}\Psi(|\bm{r}|)\{\rho(t,\bm{X}+\bm{r})-\rho(t,\bm{X})\}d\bm{r}\nonumber \\
=&\frac{1}{m}\int_{\mathbb{R}^3}\Psi(|\bm{r}|)\{r_i\frac{\partial}{\partial X_i}\rho(t,\bm{X})\nonumber \\
&+\frac{1}{2}r_{i}r_{j}\frac{\partial^2}{\partial X_i\partial X_j}\rho(t,\bm{X})+\cdots\}d\bm{r}\nonumber \\
\simeq&\frac{1}{6m}\int_{\mathbb{R}^3}\Psi(|\bm{r}|)r^{2}d\bm{r}\frac{\partial^2}{\partial X_i^2}\rho(t,\bm{X})\nonumber \\
\equiv&-\kappa\frac{\partial^2}{\partial X_i^2}\rho(t,\bm{X}).
\end{align}
Here $\kappa>0$, since $\Psi$ is attractive. The reduction from
the second to the last line is a consequence of the isotropic assumption
on $\Psi$.
This ends the derivation of the Laplacian form of $\Phi_L$.

\section{Stability analysis for case 2: $Q\ne0$\label{APP:A}}

When $Q\ne0$, the following two conditions must be satisfied simultaneously:\begin{subequations}\label{cond}
\begin{align}
 & (1+PS\alpha\lambda)SI+P\alpha\lambda Q^{2}(I-4J)=\frac{\alpha\lambda}{\pi}(1+P),\displaybreak[0]\\
 & (1+PS\alpha\lambda)(I-4J)-PS\alpha\lambda I=0.
\end{align}
\end{subequations}Because $I$ and $I-4J$ can be converted into
the Voigt functions $U$ and $V$ as\begin{subequations}\label{eq:conversion}
\begin{align}
 & \pi S^{2}I(Q,S)=U(\frac{Q}{S},\frac{1}{4S^{2}}),\\
 & \pi SQ[I(Q,S)-4J(Q,S)]=V(\frac{Q}{S},\frac{1}{4S^{2}}),
\end{align}
\end{subequations}the above conditions are recast as
\begin{align}
 & (1+PS\alpha\lambda)U+P\alpha\lambda QV=S\alpha\lambda(1+P),\displaybreak[0]\\
 & (1+PS\alpha\lambda)V-P\alpha\lambda QU=0,
\end{align}
where the Voigt functions $U$ and $V$ are defined for $x\in\mathbb{R}$
and $t>0$ as \citep{NIST}
\begin{align*}
U(x,t) & \equiv\frac{1}{\sqrt{4\pi t}}\int_{-\infty}^{\infty}\frac{1}{1+y^{2}}\exp(-\frac{(x-y)^{2}}{4t})dy,\displaybreak[0]\\
V(x,t) & \equiv\frac{1}{\sqrt{4\pi t}}\int_{-\infty}^{\infty}\frac{y}{1+y^{2}}\exp(-\frac{(x-y)^{2}}{4t})dy.
\end{align*}
Obviously both $U$ and $V$ are positive for $x>0$. Moreover $xU(x,\cdot)-V(x,\cdot)$
is positive for $x>0$, thanks to (\ref{eq:conversion}). Solving
the both conditions for $P$ tells that
\begin{equation}
\frac{S\alpha\lambda-U}{S\alpha\lambda(U+QV/S-1)}=-\frac{V}{S\alpha\lambda(V-UQ/S)},\label{eq:VU}
\end{equation}
which is solved to give
\[
S\alpha\lambda=0_{+},\quad\frac{qU^{2}+qV^{2}-V}{qU-V},
\]
where $q\equiv Q/S$ and the arguments of $U$ and $V$ are $q$ and
$1/(4S^{2})$. In order for the uniform equilibrium state to be unstable,
the second solution in the above should be larger than unity, namely
\begin{equation}
\frac{qU^{2}+qV^{2}-V}{qU-V}>1.\label{eq:hyp}
\end{equation}
We will show that this is impossible.

Because the condition \eqref{cond} is even in $Q$, we may assume
$Q>0$ (or $q>0$) without loss of generality. Because $qU-V>0$,
the condition (\ref{eq:hyp}) is reduced to
\begin{align*}
 & qU^{2}+qV^{2}-V>qU-V,\quad\text{i.e., }U^{2}+V^{2}-U>0.
\end{align*}
Now let us consider the function
\[
H(q,S)\equiv U(q,\frac{1}{4S^{2}})^{2}+V(q,\frac{1}{4S^{2}})^{2}-U(q,\frac{1}{4S^{2}}).
\]
On one hand, we have
\begin{equation}
\lim_{q\downarrow0}H(q,S)=\sqrt{\pi}SF(S)\{\sqrt{\pi}SF(S)-1\}<0,\label{eq:H(0)}
\end{equation}
because $\lim_{q\downarrow0}V(q,\frac{1}{4S^{2}})=0$ and $\lim_{q\downarrow0}U(q,\frac{1}{4S^{2}})=\frac{S}{\sqrt{\pi}}\int_{-\infty}^{\infty}\frac{1}{1+y^{2}}\exp(-S^{2}y^{2})dy=\sqrt{\pi}S\exp(S^{2})\{1-\mathrm{erf}(S)\}=\sqrt{\pi}SF(S)$.
On the other hand, because the Voigt functions are known to satisfy
the following equations:
\begin{align*}
V(x,t) & =xU(x,t)+2t\frac{\partial U(x,t)}{\partial x},\\
U(x,t) & =1-xV(x,t)-2t\frac{\partial V(x,t)}{\partial x},
\end{align*}
we have
\begin{align*}
 & \frac{\partial H}{\partial q}\\
= & 2U\frac{\partial U}{\partial q}+2V\frac{\partial V}{\partial q}-\frac{\partial U}{\partial q}\displaybreak[0]\\
= & 4US^{2}(V-qU)+4VS^{2}(1-qV-U)-(V-qU)2S^{2}\displaybreak[0]\\
= & 2S^{2}\{-2q(U^{2}+V^{2})+V+qU\}\displaybreak[0]\\
= & 2S^{2}\{-2qH-qU+V\},
\end{align*}
which is solved to yield
\begin{align*}
 & H(q,S)\\
= & H(0,S)e^{-2S^{2}q^{2}}\\
 & +\int_{0}^{q}2S^{2}\{V(r,\frac{1}{4S^{2}})-rU(r,\frac{1}{4S^{2}})\}e^{-2S^{2}(q^{2}-r^{2})}dr\displaybreak[0]\\
< & H(0,S)e^{-2S^{2}q^{2}}<0,
\end{align*}
because of $rU(r,\cdot)-V(r,\cdot)>0$ and (\ref{eq:H(0)}). Hence,
any mode with $Q\ne0$ is shown not to grow exponentially, and accordingly
the uniform equilibrium state is linear stable.


\begin{thebibliography}{99}
\bibitem{FN1}
It is rather conventional in the  kinetic theory community to use this terminology
for the limit of frequent collisions.
This limit is sometimes further classified into
the fluid-dynamic (or Navier--Stokes), diffusion (or diffusive),
and acoustic (or Euler) limits, depending on the resulting set of equations
described by the macroscopic quantities.
We follow these conventions in the present paper.
\bibitem{FNN}
The period of space domain does not affect the result and discussions for $Q\ne0$.
\end{thebibliography}
\end{document}